%% file: article_tum.tex
\documentclass[font=times]{tumarticle}

\usepackage{cite}
\usepackage{tumeicph}

\newcommand{\clsn}[1]{\textbf{#1}}

\title{mbsolve: An open-source solver tool for the Maxwell-Bloch equations}
\author[orcid=0000-0002-4030-2818, email={michael.riesch@tum.de}]{%
  Michael Riesch}
\author[orcid=0000-0003-0785-5530]{Christian Jirauschek}
\affil{Department of Electrical and Computer Engineering, Technical
  University of Munich, Arcisstr. 21, 80333 Munich, Germany}
\date{Received: 18 May 2020 / Revised: 27 March 2021 /
  Accepted: 20 May 2021 / Published (online): 16 July 2021
  \thanks{This is a post-peer-review, pre-copyedit version of an article
    published in Computer Physics Communications.
    The final authenticated version is available online at:
    \url{https://dx.doi.org/10.1016/j.cpc.2021.108097}}}

\begin{document}

\maketitle

\begin{abstract}
\input{abstract.tex}
\end{abstract}

\input{content.tex}

\bibliographystyle{osajnl}
%\bibliography{../../../Literature/literature.bib}

\input{literature.bbl}
\end{document}

%% file: abstract.tex
The Maxwell-Bloch equations are a valuable tool to model light-matter
interaction, where the application examples range from the description of
pulse propagation in two-level media to the elaborate simulation of
optoelectronic devices, such as the quantum cascade laser (QCL).
In this work, we present mbsolve, an open-source solver tool
for the Maxwell-Bloch equations. Here, we consider the one-dimensional
Maxwell's equations, which are coupled to the Lindblad equation. The
resulting generalized Maxwell-Bloch equations are treated without invoking
the rotating wave approximation (RWA). Since this full-wave treatment is
computationally intensive, we provide a flexible framework to implement
different numerical methods and/or parallelization techniques.
On this basis, we offer two solver implementations that use OpenMP for
parallelization.

%% file: content.tex
\section{Introduction}
\label{sec:introduction}

In the field of nonlinear optics, the Maxwell-Bloch equations are a valuable
tool to model light-matter interaction~\cite{boyd2008}.
Originally devised to describe the behavior of magnetic moments of nuclei
in a magnetic field~\cite{bloch1946}, the Bloch equations soon found further
application in which the dynamics of a two-level quantum mechanical system
in resonance with an optical field are
described~\cite{feynman1957,arecchi1965,abella1966}.
This version is often referred to as optical Bloch equations and, coupled
with Maxwell's equations for the optical field, was successfully applied to
model nonlinear phenomena such as self-induced
transparency~\cite{mccall1967,mccall1969}.
Later, a generalized form of the optical Bloch equations that considers an
arbitrary number of energy levels was derived~\cite{hioe1981}.
Instances of this form for three energy levels have been used e.g., to
describe electromagnetically induced transparency and the
related slow light propagation~\cite{hau1999,liu2001}.
Similarly, further application examples of this generalized form can be
related to the propagation of light in different media, and considering quantum
mechanical systems with two~\cite{ziolkowski1995,cartar2017,gladysz2020},
three~\cite{slavcheva2002,sukharev2011} or
six~\cite{marskar2011} energy levels, respectively.

While in earlier studies the media considered were mostly of gaseous form,
advances in nanotechnology paved the way for solid state optoelectronic
devices that exhibit (at least partially) coherent light-matter interaction,
as is adequately described by the Maxwell-Bloch
equations~\cite{jirauschek2019ats}.
Here, the quantum cascade laser (QCL)~\cite{kazarinov1971,faist1994} is a
notable example.
In many studies (e.g., \cite{wang2007,menyuk2009,choi2010,gkortsas2010,freeman2013,jirauschek2014,talukder2014,wang2015active,tzenov2016,vukovic2017,tzenov2018,columbo2018}),
its dynamical behavior was simulated in this framework.
Also quantum dot devices have been extensively modeled based on Maxwell-Bloch
equations~\cite{nielsen2007,majer2010,cartar2017,slavcheva2019}.
For a detailed overview of the applications, the reader is referred
to our recent review paper~\cite{jirauschek2019ats}.

In the majority of related studies, the rotating wave approximation (RWA)
in combination with the slowly varying envelope approximation (SVEA) is
invoked.
Naturally, these approximations may omit certain features of the
solution~\cite{ziolkowski1995}.
We assume that those features are crucial in the scope of simulations of
quantum cascade laser frequency
combs~\cite{riesch2018atrasc1,riesch2018iqclsw}, which is the main
application of the Maxwell-Bloch equations in our research.
Hence, we aim to avoid the RWA and SVEA.

It should be noted that even when those approximations are used, analytical
solutions are not generally available.
As a consequence, we need to resort to numerical methods and solver tools
that implement them.
Although several numerical approaches for the Maxwell-Bloch equations have
been discussed and compared in related literature~\cite{ziolkowski1995,bidegaray2001,bidegaray2003,sukharev2011,jirauschek2019ats,marskar2011,saut2006,deinega2014,riesch2018oqel,riesch2019cleo1,riesch2020nusod1}, there is no definitive statement
on what the best approach is.
Also, just as with many problems in science and engineering, it is improbable
that a single numerical method can cover all use cases.
For example, while invoking the RWA/SVEA may not be suitable for quantum
cascade laser frequency combs, it is perfectly reasonable for a multitude
of problems.
Therefore, the solver tool should provide support for multiple numerical
methods in order to evaluate and compare different approaches.
Here, a small and flexible code base is beneficial for rapid prototyping.

At the same time, the tool should permit productive usage with established
numerical methods in the scope of our research, which leads to the following
requirements.
As already stated, a full-wave treatment of the optical field
is desired (i.e., the rotating wave approximation should not be invoked).
Secondly, as the full-wave treatment is computationally more intensive, the
resulting numerical operations should be efficiently executed in parallel.
Then, the solver should be able to deal with multiple sections of different
materials, flexible initial and boundary conditions, source terms, and an
arbitrary number of energy levels.
Finally, the source code of the solver should be publicly available in order
to allow extensions of the software and to improve the reproducibility of
simulation results.

Unfortunately, the majority of the solver tools used in related work are
not publicly available (e.g., in~\cite{ziolkowski1995,bidegaray2001,bidegaray2003,sukharev2011,marskar2011,saut2006}).
In the following we discuss the few exceptions to that rule.
For example, the Electromagnetic Template Library (EMTL)~\cite{emtl}
is a free C++ library with Message Passing Interface (MPI) support and has
been used e.g., to model single quantum emitters~\cite{deinega2014}.
However, it is only available in binary form which makes it impossible to
extend the library and port it to new computing architectures.
The Freetwm~\cite{freetwm} project is an open-source MATLAB code that
solves the 1D Maxwell-Bloch equations.
But to the best of our knowledge, it uses the rotating wave approximation and
does not support alternative numerical methods.
Also, solvers written and executed in MATLAB typically feature inferior
performance than implementations in compiled languages such as C++.
Finally, the open-source project MEEP~\cite{oskooi2010} is a fully versed
finite-difference time-domain (FDTD) solver for Maxwell's equations with
parallelization support (using MPI) and a flexible user interface.
Also, it features support for multi-level quantum mechanical systems.
It is widely accepted in the simulation community and, therefore, it is a
promising project to base our future simulations onto.
But one has to acknowledge that the evaluation and comparison of different
numerical methods for the Maxwell-Bloch equations is hardly possible with
such a large code base.

With this overview of existing approaches in mind, we decided to start
developing our own project named mbsolve.
The minimal and flexible code base already allowed comparisons of different
numerical methods~\cite{riesch2018oqel,riesch2019cleo1,riesch2020nusod1}, as
well as a discussion of their efficiency on parallel
architectures~\cite{riesch2019pasc}.
Subsequently, the mbsolve project has been used productively in an
investigation of harmonic mode locking in terahertz quantum cascade
lasers~\cite{wang2020ultrafast} and the modeling of terahertz frequency
combs interacting with a graphene-coated reflector~\cite{mezzapesa2020}.
However, it has not yet been introduced to potential users who wish to use
this simulation software for their applications.
In the work at hand, we would like to make up for this and present the
mbsolve project.

In the following section, we introduce the Maxwell-Bloch equations as they
are used in the scope of the mbsolve software, and discuss the different
numerical methods that are employed to solve them.
In Section \ref{sec:implementation}, we present the implementation of
mbsolve and discuss in detail the fundamental components.
Namely, those are the basic library that provides the description of a
simulation setup, an implementation of a solver using the OpenMP standard
for parallelization, and a library that exports simulation results into
the HDF5 format.
In Section~\ref{sec:applications}, we verify the correct operation of our
implementation and discuss the features of the mbsolve software with the
help of four application examples from related literature.
Finally, after a short summary we conclude in Section~\ref{sec:conclusion}
with an outlook on possible extensions of the software.

\section{The Maxwell-Bloch equations}
\label{sec:mb}

As already outlined, the Maxwell-Bloch equations describe the interaction of
an electromagnetic field with quantum mechanical systems. In the most general
picture, the electric field $\vec E(\vec r, t)$ and the magnetic field
$\vec H(\vec r, t)$ depend on the three-dimensional space coordinate $\vec r$
and time $t$.
The quantum mechanical systems are assumed to be uniformly distributed over
space and are generally represented by the density matrix
$\hat \rho(\vec r, t)$. While the electromagnetic field is treated
classically using Maxwell's equations (in 3D), the density matrix is a concept
from quantum mechanics. Its time evolution is governed by a master equation
of the form
\begin{equation}
  \partial_t \hat{\rho} = -\mathrm{i}\hbar^{-1} \left[ \hat{H}, \hat{\rho}
    \right] + \mathcal{D}(\hat{\rho}),
  \label{eq:master}
\end{equation}
where $\hbar$ is the reduced Planck constant and $[ \cdot, \cdot]$ denotes
the commutator. The Hamiltonian $\hat H = \hat H_0 - \hat{\vec\mu} \vec E$
consists of the static Hamiltonian $\hat H_0$ and the interaction term
$\hat{\vec\mu} \vec E$, where $\hat{\vec\mu}$ is the dipole moment
operator. The superoperator $\mathcal{D}(\hat\rho)$ accounts for the
non-unitary contribution to the time evolution of the density matrix. Its
form also specifies the type of master equation. For example, if the
superoperator is given in Lindblad form, Eq.~(\ref{eq:master}) is referred to
as Lindblad equation. At this point, the quantum mechanical systems are
controlled by the electromagnetic field. The interaction cycle is completed
by the introduction of a polarization term
\begin{equation}
  \partial_t \vec P =
  n_\mathrm{3D} \trace\{ \hat{\vec\mu} \partial_t \hat \rho\}
\end{equation}
that enters Ampere's law in Maxwell's equations.
Here, $n_\mathrm{3D}$ is the density of quantum mechanical systems, and
$\trace\{\}$ is the trace operation~\cite{jirauschek2019ats}.

\subsection{Generalized Maxwell-Bloch equations in 1D}

The resulting three-dimensional form of the Maxwell-Bloch equations is quite
complex. In the scope of this work we address a simplified version that is
relevant for two classes of applications. The first class consists of
simulation models in which the plane wave approximation~\cite{siegman1986}
is a reasonable assumption. Those models can be found frequently in related
literature (e.g.,~\cite{ziolkowski1995,marskar2011,song2005}). The second
class of applications contains the simulation of optoelectronic devices with
a waveguide geometry that permits the separation of
transversal and longitudinal modes. As a consequence, only the propagation
direction has to be considered in the Maxwell-Bloch equations, whereas the
change in transversal directions is accounted for using an effective
refractive index. A detailed description of this procedure is given
in~\cite{jirauschek2019ats}.

The simplified version of Maxwell-Bloch equations can be written in
terms of the field components $E_z(x, t)$ and $H_y(x, t)$, where $x$ is the
propagation direction, and $y$, $z$ are the transversal coordinates.
Then, Maxwell's equations can be reduced to equations for the time evolution
of the electric field
\begin{equation}
  \partial_t E_z = \epsilon^{-1} \left( -\sigma E_z - \Gamma \partial_t P_z +
  \partial_x H_y \right)
  \label{eq:efield}
\end{equation}
and the magnetic field
\begin{equation}
  \partial_t H_y = \mu^{-1} \partial_x E_z,
  \label{eq:hfield}
\end{equation}
where the conductivity $\sigma$, the permittivity
$\epsilon = \epsilon_0 \epsilon_\mathrm{r}$, and the permeability
$\mu = \mu_0 \mu_\mathrm{r}$
can be related to a (generally complex) effective refractive index.
Additionally, the overlap factor $\Gamma$ is introduced, which accounts
for a partial overlap of the transversal mode with the quantum mechanical
systems and can be set to unity in cases where it is not required.
The effective refractive index and the overlap factor are commonly used
in optoelectronic device simulations and take into account the properties of
the bulk material(s) in the setup as well as the waveguide geometry
(if any).
Generally, they depend on the position $x$ and the frequency, which applies
here for the material parameters $\epsilon$, $\sigma$, $\mu$ and $\Gamma$.
We note that the frequency dependence reflects the chromatic dispersion of the
background medium, which combines waveguide and bulk dispersion.
In the work at hand, we ignore this frequency dependence
but note that the inclusion of the background
dispersion should be in the focus of future work.
The dispersion that stems from the quantum mechanical systems, on the other
hand, is included in the polarization $\vec P$ and will be considered.

The variation of the material parameters in propagation direction is usually
piecewise constant, in which case different regions of materials can be used
to address this dependence. For each region, Eqs.~(\ref{eq:efield}) and
(\ref{eq:hfield}) describe the electromagnetic field, where the material
parameters are then constants.
Within this model, the quantum mechanical systems are distributed only in
the propagation direction and can be represented by the density matrix
$\hat \rho(x, t)$.
Furthermore, the dipole moment operator is now assumed to have only one
non-zero component $\hat\mu_z$~\cite{jirauschek2019ats}. Therefore, the
polarization contribution that stems from the quantum mechanical systems can
be written as
\begin{equation}
  \partial_t P_z =
  n_\mathrm{3D} \trace\{ \hat{\mu}_z \partial_t \hat \rho\},
\end{equation}
and the time evolution of the density matrix is described by the master
equation
\begin{equation}
  \partial_t \hat{\rho} = -\mathrm{i}\hbar^{-1}
  \left[ \hat{H}_0 - \hat \mu_z E_z, \hat{\rho}
    \right] + \mathcal{D}(\hat{\rho}).
  \label{eq:master1D}
\end{equation}

Now, as the system of partial differential equations is introduced, it is
necessary to specify the initial and boundary conditions. The electric and
magnetic fields require initial values $E(x, t=0)$ and $H(x, t=0)$,
respectively. While many setups simply assume both to be zero, the fields
may be initialized with random values to model spontaneous emission. The
latter is typically applied during the simulation of lasers.
As to boundary conditions, it is generally assumed that the electromagnetic
wave is reflected with a certain reflectivity $R$ at the simulation domain
boundary. This assumption covers the case of perfect ($R=1$) and
semi-transparent ($0 < R < 1$) mirrors, which are often considered in
optoelectronic devices, as well as perfectly matched layer (PML) boundary
conditions ($R=0$). It should be noted that the reflectivity values of the
two boundaries $R_1 \neq R_2$ may be different. Since the master
equation~(\ref{eq:master1D}) does not include a spatial derivative, only the
initial value of the density matrix at each point is required, such as
thermal equilibrium (lowest energy level has the largest population) or
inversion (some higher energy level, the so-called upper laser level, has the
largest population).
Finally, we note that source terms are often included in related literature.
For example, an incoming electromagnetic pulse of Gaussian or sech shape is
modeled by a source term in the electric field.

Together with those initial and boundary conditions, the
differential equations (\ref{eq:efield})-(\ref{eq:master1D}) form the
generalized Maxwell-Bloch equations in 1D. If we reduce the model further by
ignoring the propagation effects altogether, the equation system is nothing
more than the master equation~(\ref{eq:master1D}) that describes the time
evolution for a single quantum mechanical system. As a consequence, the
software project presented in the work at hand may also serve as solver for
e.g., the Lindblad equation. Starting again from the one-dimensional version,
we note that the generalized Maxwell-Bloch equations are able to treat an
arbitrary number of energy levels $N$. The original Maxwell-Bloch equations
can be derived by setting $N=2$. At this point, the rotating wave
approximation (RWA) is usually applied in order to reduce further the
complexity of the equation system. However, we shall not go along this way,
but discuss numerical methods that are able to solve the one-dimensional
generalized Maxwell-Bloch equations.

\subsection{Numerical treatment}

Here, we can divide the discussion in three parts.
The first part is dedicated to numerical methods that solve Maxwell's
equations.
As already stated above, we focus on methods that solve the full wave
equations, i.e., that do not invoke the rotating wave or slowly varying
envelope approximation.
Another part deals with the numerical treatment of the master
equation~(\ref{eq:master1D}).
In between, the coupling of both parts is discussed.
Throughout this subsection, we base on our recent review paper, in which
the state of the art in numerical methods for the Maxwell-Bloch
equations is reviewed in more detail~\cite{jirauschek2019ats}.

The majority of related work uses on the finite-difference time-domain
(FDTD) method to solve Maxwell's equations.
Clearly, it is one of the standard approaches in computational
electromagnetics, and its simplicity allows straightforward implementation
of source terms and sharp material boundaries.
As disadvantages, its numerical dispersion and the need for a constant
spatial discretization must be mentioned.
The usual remedy for the numerical dispersion is to decrease the spatial and
temporal grid spacing, which in turn increases the computational workload.
As an alternative, the pseudo-spectral time-domain (PSTD) method has been
used in related work~\cite{saut2006,marskar2011}.
The PSTD method calculates the spatial derivatives accurately in a
pseudo-spectral domain, in which they are reduced to mere multiplication
operations.
Thereby, the spatial contribution of the numerical dispersion is eliminated
and the temporal contribution is significantly reduced.
However, this comes at the cost of potentially expensive calls to the fast
Fourier transform (FFT), and complex implementations of boundary conditions
and source terms.
The need of the FDTD method for a constant spatial discretization can lead to
inefficient discretization patterns in setups with different media.
The grid spacing must be chosen to suit the medium with the largest
effective refractive index and will be unnecessarily small for media with
smaller refractive indices.
Numerical methods that adapt the spatial discretization variably to the problem
are used to solve Maxwell's equations, but to the best of our knowledge they
have not been applied in the scope of the Maxwell-Bloch equations.
In this work, we follow the majority of related literature and focus on the
FDTD method.
Nevertheless, the resulting software should be designed that it
can serve as basis for implementations of alternative numerical methods, such
as the PSTD method or methods that use adaptive spatial grids.

\begin{figure}
  \centering
  \input{figures/tikz_fdtd_os.tex}
  \caption{The standard Yee grid of the FDTD method combined with the
    discretization of the density matrix with respect to time and space.
    Electric and
    magnetic field are denoted with orange crosses and blue circles,
    respectively. The density matrix discretization is marked using green
    squares. The arrows indicate the data dependencies during the update of
    three quantities.
    Reprinted from Riesch M., et al.~\cite{riesch2019pasc} (CC BY 4.0).}
  \label{fig:fdtd}
\end{figure}
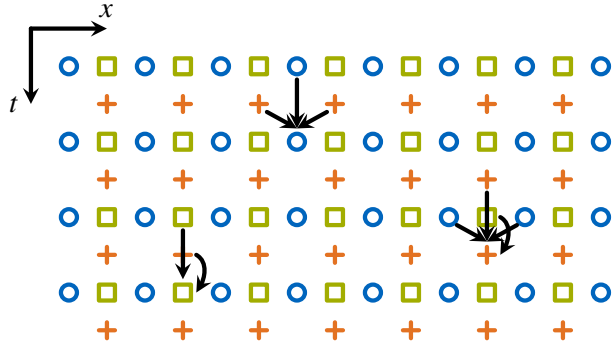

Figure~\ref{fig:fdtd} shows the standard Yee grid of the FDTD method, i.e.,
the discretization of the electric field $E_z$ and magnetic field $H_y$ as
function of the spatial index $m$ and the temporal index $n$. The main
feature of the Yee grid is that the electric and magnetic field are staggered
in time and space.
This clever choice leads to the central difference
equations~\cite{taflove2005}
\begin{equation}
  \begin{aligned}
    \frac{E_z^{m, n+1} - E_z^{m, n}}{\Delta t} =
    &- \epsilon^{-1}\sigma \frac{E_z^{m,n+1} + E_z^{m, n}}{2}
    - \epsilon^{-1} \Gamma \left(\partial_t P_z\right)^{m, n + 1/2}\\
    &+ \epsilon^{-1}
    \frac{H_y^{m + 1/2,n + 1/2} - H_y^{m - 1/2,n + 1/2}}{\Delta x}
  \end{aligned}
  \label{eq:ediscrete}
\end{equation}
and
\begin{equation}
  \frac{H_y^{m + 1/2,n + 1/2} - H_y^{m + 1/2,n-1/2}}{\Delta t} = \mu^{-1}
  \frac{E_z^{m + 1, n} - E_z^{m, n}}{\Delta x}
  \label{eq:hdiscrete}
\end{equation}
for Eqs.~(\ref{eq:efield}) and (\ref{eq:hfield}), respectively. Here,
$\Delta x$ is the spatial grid point size, and $\Delta t$ is the temporal
discretization size.
As to the discretization of the density matrix, there are two choices, which
are referred to as weak and strong coupling,
respectively~\cite{bidegaray2003}. The difference
between those choices is the temporal discretization, where strong coupling
uses the same discretization points for both the electric field and the
density matrix. Weak coupling, as depicted in Fig.~\ref{fig:fdtd}, uses
different discretization points, which are staggered in time.

Regardless of the coupling, the remaining task is the update step of the
density matrix from one time value to the next one. This operation can be
formally written as
\begin{equation}
  \hat \rho_{n+1/2} = \exp(\mathcal{L}\Delta t) \hat \rho_{n-1/2}
  \approx
  \mathcal{U}_n(\hat \rho_{n-1/2}),
\end{equation}
where the Liouvillian superoperator $\mathcal{L}$ represents the right-hand
side of Eq.~(\ref{eq:master1D}). The solution of Eq.~(\ref{eq:master1D}) for
a certain time step $\Delta t$ can be specified with the help of the
exponential superoperator.
In practice, however, this formal expression can only be approximated using
a certain update operation $\mathcal{U}_n$, which serves as placeholder
for different numerical methods.
Indeed, several candidates with different accuracies and complexities have
been suggested in related literature, which can be roughly divided into
three groups~\cite{jirauschek2019ats,riesch2019jcomp}.
Several studies use the Crank-Nicolson method to solve the master equation,
where the implicit nature of this method is usually resolved using a
predictor-corrector approach~\cite{ziolkowski1995,slavcheva2002}.
However, it has been pointed out that neither the Crank-Nicolson method nor the
predictor-corrector approach guarantees the preservation of the positivity
of the density matrix~\cite{bidegaray2001,riesch2019jcomp}.
Alternatively, higher-order Runge-Kutta methods have been employed
(e.g., in~\cite{sukharev2011,cartar2017}).
Although they typically feature higher accuracy than the first group of
methods, there is still no guarantee that the positivity of the density
matrix is preserved.
The third group of approaches revolves around solving the master equation
exactly for one time step, which is related to calculating matrix
exponentials.
Since this operation is very costly, operator splitting techniques are often
applied~\cite{bidegaray2001,bidegaray2003,marskar2011}.
This group of methods is the one with the highest computational complexity but
guarantees the preservation of the density matrix properties.
For a thorough review of numerical methods for the master equation the
reader is referred to our recent
reviews~\cite{jirauschek2019ats, riesch2019jcomp}.
We would like to point out, however, that there is still no definite answer
on what the best numerical method is.
This question has been in the focus of recent
work~\cite{riesch2018oqel,riesch2019cleo1,riesch2020nusod1} and will
be discussed elsewhere in more detail.

In mbsolve, the established operator splitting method presented
in~\cite{bidegaray2001} is implemented for productive usage and as reference
for promising new methods.
This method guarantees to preserve the properties of the density matrix and
is therefore well-suited for long-term simulations, which are often required
in the scope of the simulation of optoelectronic devices.
Additionally, there is an adapted form of the operator splitting
method~\cite{riesch2017c} that is currently evaluated.
As we have already pointed out, one of the primary design goals of mbsolve is
the
support of different numerical methods, which naturally includes different
approaches to solve the master equation.
Thereby, mbsolve will aid the comparison between various promising candidates.

\section{Implementation of mbsolve}
\label{sec:implementation}

In this section, we present the architecture and the implementation details
of the mbsolve software.
The modular architecture was designed with the requirements from
Section~\ref{sec:introduction} in mind, which we revisit briefly in the
following.
As we already stated, it is crucial that mbsolve supports different
numerical methods for the Maxwell-Bloch equations.
We note that a flexible way to combine methods for Maxwell's equations and
methods for the master equation would be beneficial.
Then, for example, it would not be required to implement the FDTD method
multiple times when evaluating different algorithms to solve the master
equation.
Similarly, there are different parallelization techniques (OpenMP
for shared memory systems, MPI for distributed memory systems, CUDA for
NVIDIA graphics processing units (GPU), etc.), and mbsolve should be able
to handle different techniques.
Thereby, available simulation hardware can be targeted and exploited.
A preliminary study found that the Maxwell-Bloch equations can be
efficiently solved on GPUs~\cite{riesch2017cleo}, but the software should
also work on a regular desktop PC without a high-end graphics
card.
As already mentioned, we found that a small, yet flexible and
extensible, code base is beneficial in order to achieve this.
Finally, we envisage to share the resulting software project with the
scientific community.
Here, several measures must be taken to guarantee that other researchers
can acquire, install and use the software~\cite{riesch2020pone}.
While we discuss those measures in more detail below, the fundamental
decision is the choice of programming language.
We decided to use the C++ programming language for performance reasons,
and offer bindings for Python in order to provide an easy-to-use interface
for the researchers.
Both programming languages are established in the scientific community and
should constitute a reasonable choice~\cite{riesch2020pone}.

\begin{figure}
  \centering
  \includegraphics[width=0.5\textwidth]{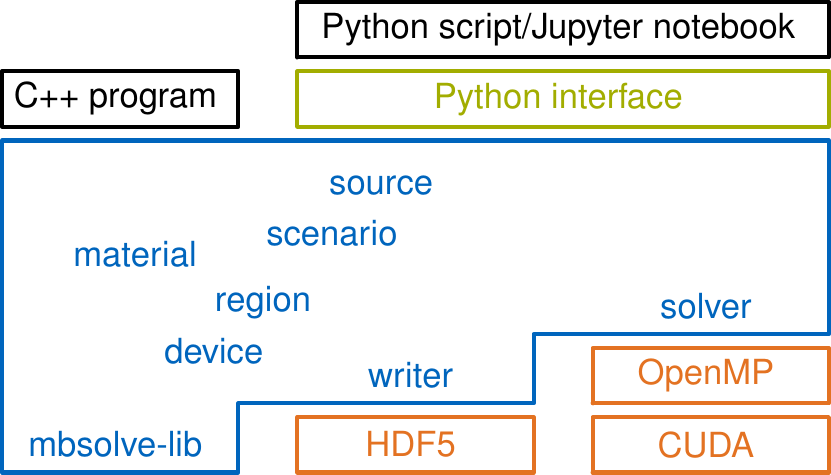}
  \caption{Overview of the mbsolve project.
    Reprinted from M. Riesch, The QCL Stock
    Image Project~\cite{qclsip} (CC BY 4.0).}
  \label{fig:mbsolve_overview}
\end{figure}

During the implementation of the mbsolve software, the required flexibility
was guaranteed by modularization and clearly defined interfaces. Those
design criteria lead to the architecture depicted in
Fig.~\ref{fig:mbsolve_overview}. The mbsolve-lib base library constitutes the
fundamental part of the software, as it provides an object oriented framework
to define a simulation setup as well as the infrastructure to add solver
and writer components.
As the name suggests, the solver components implement numerical
methods that solve the specified simulation setup using different
parallelization techniques. After the solver has completed its work, the
writer component is responsible for writing the simulation results into a
file. In principle, writers for any file format can be implemented. However,
we found that the Hierarchical Data Format (HDF) fulfills all of our needs.

In the following, we present in more detail the mbsolve-lib base library, two
exemplary solvers based on the OpenMP standard and the numerical methods
described in~\cite{bidegaray2001,riesch2017c}, respectively, and the writer
for the HDF5 file format.
Subsequently, we discuss the measures taken to create a sustainable
open-source software project out of mbsolve.

\subsection{The mbsolve-lib base library}

The object oriented framework that describes a simulation setup can be
divided into two parts. The device part contains all properties of a setup
that are static. These properties include, for example, the composition of
the device under simulation in terms of regions and materials. The dynamic
properties, on the other hand, are grouped into a scenario. For instance,
simulation properties, such as the number of grid points used, or source
terms are defined in the scenario. Thereby, it is possible to simulate the
same device under different conditions, which was used, e.g., during the
investigation of seeding effects in a quantum cascade
laser~\cite{freeman2013}. After device and scenario are defined, the user
can pass them to a solver, which calculates the simulation results, and
optionally to a writer, which exports the results to a file.

\subsubsection{Device setup and boundary conditions}

The device is represented by a class of the same name.
It contains a collection of \clsn{region} objects, which models a section
of the device in which the material parameters are constant.
A region is defined by the $x_\mathrm{start}$ and $x_\mathrm{end}$ coordinates
as well as a pointer to a certain material.
The envisaged materials have to be present in a static collection when the
region is created.
Thereby, material parameters are not directly stored in the regions.
This enables the efficient treatment of periodic structures, where only few
materials are repeatedly used in many different regions.

Each material is an instance of the eponymous class, which contains
the electromagnetic properties as well as the description of the quantum
mechanical systems.
The former consist of the permittivity $\epsilon$, the  permeability $\mu$,
the overlap factor $\Gamma$, and the linear loss term
\begin{equation}
  \alpha_0 = \frac{\sigma}{2\epsilon c}
  = \sqrt{\frac{\mu}{\epsilon}} \frac{\sigma}{2},
\end{equation}
which is related to the conductivity $\sigma$~\cite{siegman1986}.
Here, $c = (\mu\epsilon)^{-1/2}$ is the speed of light in the material.
The latter is incorporated as pointer to the class \clsn{qm\_description},
which represents a quantum mechanical system.
Here, the usage of a pointer allows polymorphism, i.e., the quantum
mechanical description can assume different forms.

In the general form, this description requires the density of quantum
mechanical systems $n_\mathrm{3D}$, the Hamiltonian $\hat H_0$ and the dipole
moment operator $\hat \mu_z$, as well as the superoperator $\mathcal{D}$.
For $\hat H_0$ and $\hat \mu_z$ the class \clsn{qm\_operator} can be used.
Since quantum mechanical operators can be expressed as Hermitian matrices,
it is sufficient to store the real entries of the main diagonal as well as
the complex entries of the upper triangle of the matrix.
For example, a diagonal Hamiltonian
\begin{equation}
  \hat H_0 = \begin{bmatrix} \omega_1 & 0 & 0\\ 0 & \omega_2 & 0\\
    0 & 0 & \omega_3 \end{bmatrix}
\end{equation}
can be created from a real vector $[ \omega_1, \omega_2, \omega_3]$.
In this case it is not necessary to provide a complex vector for the
off-diagonal elements since they are set to zero by default.
On the other hand, for a dipole moment operator
\begin{equation}
  \hat \mu_z =
  \begin{bmatrix}
    0 & \mu_{12} & \mu_{13}\\
    \mu_{21} & 0 & \mu_{23}\\
    \mu_{31} & \mu_{32} & 0
  \end{bmatrix}
\end{equation}
the main diagonal elements are defined by the real vector $[0, 0, 0]$ and
the off-diagonal elements are represented by the complex vector
$[ \mu_{12}, \mu_{13}, \mu_{23} ]$.

The treatment of superoperators is more complex, since the action of the
superoperator on a quantum mechanical operator must be determined.
Also, different choices of the superoperator are reasonable.
Therefore, we implemented the class \clsn{qm\_superoperator} and derived
a sub-class \clsn{qm\_lindblad\_relaxation} that represents the resulting
Lindblad superoperator.
This approach allows future extensions, yet covers all current
application examples, since the Lindblad form of the master equation is the
most general Markovian description~\cite{breuer2002}.
The Lindblad superoperator~\cite{lindblad1976,gorini1976} can be written as
\begin{equation}
  \mathcal{D}(\hat \rho) = \sum_{i,j=1}^{N^2 - 1}
  c_{ij}
  \left[
    \hat F_i \hat \rho \hat F_j^\dagger - \frac{1}{2}
    \left(
    \hat F_j^\dagger \hat F_i \hat \rho + \hat \rho \hat F_j^\dagger \hat F_i
    \right)
    \right],
  \label{eq:lindbladdiss}
\end{equation}
where $N$ is the dimension of the underlying Hilbert space $\mathcal{H}$,
i.e., the number of energy levels under consideration.
Furthermore, the coefficients $c_{ij}$ form a positive semi-definite matrix $C$
and the traceless operators $\hat F_i$ constitute an orthonormal basis of
the space $B(\mathcal{H})$ of all bounded operators on the Hilbert space
$\mathcal{H}$ with the restriction that $\hat F_{N^2}$ is proportional to the
identity operator.
It should be noted that the choice of the operators $\hat F_i$ is not unique,
as the resulting dynamical behavior is invariant under certain transforms,
e.g., any unitary transform~\cite{breuer2002}.
Preferably, we choose a set of operators that allows a straightforward
physical interpretation of the dissipation superoperator.
Such a suitable choice is presented in \cite{schirmer2004} and consists of
$N(N-1)$ operators of the form $\hat F_{k} = \ket{i}\bra{j}$,
where the indices $i,j \in [1;N], i \neq j$ are mapped to the index
$k \in [ N; N^2 - 1]$, and $N-1$ diagonal operators of the form
\begin{equation}
  \hat F_k = \frac{1}{\sqrt{k(k+1)}}
  \left( \sum_{s=1}^k \ket{s}\bra{s} - k \ket{k+1}\bra{k+1} \right),
  \label{eq:lindbladops}
\end{equation}
where $k \in [1;N-1]$.
The latter group of operators can be related to the Pauli matrices and
Gell-Mann matrices (for $N = 2$ and $N = 3$, respectively).
By plugging the operator expressions into Eq.~(\ref{eq:lindbladdiss}), it
becomes apparent that the time evolution of the populations is determined
by the operators $\hat L_k, k \in [ N; N^2 - 1]$ alone, which are in the
following referred to as relaxation operators.
As the name suggests, each operator represents a relaxation or scattering
process from an energy level $\ket{i}$ to another level $\ket{j}$ with the
scattering rate $\gamma_{ji, i \neq j}$, where the mapping between the indices
$i, j$ and $k$ mentioned before has been reverted.
Assuming that the energy levels are sorted by their corresponding energy
value in descending order, the scattering processes can be
divided into forward scattering ($i < j$, e.g., spontaneous emission) and
backscattering ($j < i$, e.g., optical pumping) processes.
By relating the coefficients $c_{ij}$ to the scattering rates $\gamma_{ij}$
in a suitable fashion, a clear correspondence between mathematical description
and physical processes can be established~\cite{jirauschek2019ats}.
As a result, the non-unitary time evolution of the populations $\rho_{ii}$ is
governed by
\begin{equation}
  \left. \partial_t
    \begin{bmatrix} \rho_{11}\\ \rho_{22}\\ \vdots \\ \rho_{NN} \end{bmatrix}
    \right|_\mathcal{D} =
    \begin{bmatrix}
      -\tau_1^{-1} & \gamma_{12} & \dots & \gamma_{1N}\\
      \gamma_{21} & -\tau_2^{-1} & \dots & \gamma_{2N}\\
      \vdots & \vdots & \ddots& \vdots\\
      \gamma_{N1} & \gamma_{N2}& \dots & -\tau_N^{-1}
    \end{bmatrix}
    \begin{bmatrix} \rho_{11}\\ \rho_{22}\\ \vdots \\ \rho_{NN} \end{bmatrix},
    \label{eq:relax_pop}
\end{equation}
where the inverse population life times
$\tau_j^{-1} = \sum_{i \neq j} \gamma_{ij}$ ensure that the trace
of the density matrix is preserved.
The time evolution of the off-diagonal elements $\rho_{ij, i \neq j}$ of the
density matrix (coherence terms), on the other hand, is affected by all
operators $\hat F_k, k \in [1, N^2 - 1]$.
This process, commonly referred to as dephasing, can be described as
\begin{equation}
  \left. \partial_t \rho_{ij} \right|_\mathcal{D} =
  - \left[
    \frac{1}{2}
    \left( \tau_i^{-1} + \tau_j^{-1} \right)
    + \gamma_{ij,\textrm{p}}
    \right]
  \rho_{ij},  \quad i \neq j,
\end{equation}
where we distinguish between the life time contribution to dephasing, and
the pure dephasing contribution with the rate $\gamma_{ij,\textrm{p}}$.
The latter can be expressed as
\begin{equation}
  \gamma_{ij,\textrm{p}} =
  \frac{1}{2} \sum_{m,n=1}^{N-1} c_{mn} \left(F_{m, ii} - F_{n, jj}\right)^2,
  \label{eq:puredeph}
\end{equation}
where the $F_{k, ii}$ denote the matrix elements of the operators
$\hat F_k$~\cite{oi2012}.
The concept of pure dephasing rates is commonly used in theoretical and
experimental work, where the rates are usually considered free parameters
that can be determined experimentally, calculated from microscopic models,
or chosen phenomenologically~\cite{jirauschek2019ats}.
However, as can be seen from Eq.~(\ref{eq:puredeph}) there are constraints
on the dephasing rates since the coefficients $c_{mn}$ must form a positive
semi-definite matrix~\cite{oi2012}.
In order to provide a physically accurate, yet practice oriented interface,
we offer a user-friendly constructor for \clsn{qm\_lindblad\_relaxation}.
It accepts a matrix
\begin{equation}
  R =
  \begin{bmatrix}
      0 & \gamma_{12} & \dots & \gamma_{1N}\\
      \gamma_{21} & 0 & \dots & \gamma_{2N}\\
      \vdots & \vdots & \ddots& \vdots\\
      \gamma_{N1} & \gamma_{N2}& \dots & 0
  \end{bmatrix},
\end{equation}
which is similar to the matrix introduced in Eq.~(\ref{eq:relax_pop}), but
ignores the elements on the main diagonal.
Those elements represent the inverse population life times and can be
determined based on the off-diagonal elements of $R$.
As second parameter, the constructor of \clsn{qm\_lindblad\_relaxation}
accepts a $N(N-1)/2$ real vector $[ \gamma_{12,\textrm{p}}, \gamma_{13,\textrm{p}},
  \gamma_{23,\textrm{p}},  \gamma_{14,\textrm{p}},  \gamma_{24,\textrm{p}}, \dots,
  \gamma_{N-1,N,\textrm{p}} ]$ with the pure dephasing rates, where the same
ordering of off-diagonal elements as in \clsn{qm\_operator} is used.
The constructor tries to convert the dephasing rates into a corresponding
coefficient matrix $C$ and checks whether this matrix is positive
semi-definite.
If the conversion or the check fails, a warning is emitted.

For the original Maxwell-Bloch equations, however, the general quantum
mechanical description is unnecessarily complex.
This form considers only two energy levels and a diagonal Hamiltonian
$\hat H_0$, and usually neglects the static dipole moments
($\mu_{z,22} - \mu_{z,11} \approx 0$).
For two energy levels, the master equation (\ref{eq:master}) reads
\begin{equation}
  \begin{aligned}
    \partial_t
    \begin{bmatrix}
      \rho_{11} & \rho_{12}\\
      \rho_{21} & \rho_{22}
    \end{bmatrix} = &
    - \mathrm{i} \hbar^{-1} \left[
      \begin{bmatrix}
        H_{0,11} - \mu_{z, 11} E & - \mu_{z, 12} E\\
        - \mu_{z, 21} E & H_{0,22} - \mu_{z, 22} E
      \end{bmatrix},
      \begin{bmatrix}
        \rho_{11} & \rho_{12}\\
        \rho_{21} & \rho_{22}
      \end{bmatrix}
      \right]\\
    & +
    \begin{bmatrix}
      - \gamma_{21} \rho_{11} + \gamma_{12} \rho_{22} &
      - \left[ \frac{1}{2}
        \left( \gamma_{12} + \gamma_{21} \right) + \gamma_{12,\mathrm{p}} \right]
      \rho_{12}\\
      - \left[ \frac{1}{2}
        \left( \gamma_{12} + \gamma_{21} \right) + \gamma_{21,\mathrm{p}} \right]
      \rho_{21} &
      \gamma_{21} \rho_{11} - \gamma_{12} \rho_{22}
    \end{bmatrix}.
  \end{aligned}
  \label{eq:master2lvl}
\end{equation}
Additionally, there is the constraint $\rho_{11} + \rho_{22} = 1$ on the
populations, and for the coherence terms $\rho_{12} = \rho_{21}^*$ holds.
Therefore, it is sufficient to determine the population inversion
$w = \rho_{22} - \rho_{11}$, as the populations $\rho_{11} = (1 - w)/2$  and
$\rho_{22} = (1 + w)/2$ can be derived from this quantity, and one of the
coherence terms (usually $\rho_{21}$)~\cite{jirauschek2019ats}.
Following this approach, and by assuming that the dipole moment $\mu_{z,21}$
is real, Eq.~(\ref{eq:master2lvl}) can be brought into the form
\begin{subequations}
\begin{align}
  \partial_t \rho_{21} &=
    - \mathrm{i} \omega_{21} \rho_{21}
    - \mathrm{i} w \Omega_\mathrm{R}
    - \gamma_{2} \rho_{21},\\
  \partial_t w &=
      4 \Omega_R \Im \left\{ \rho_{21} \right\}
    - \gamma_1 \left( w - w_0 \right),
\end{align}
\end{subequations}
which are the optical Bloch equations in the original form.
Here, $\omega_{21} = \hbar^{-1} ( H_{0,22} - H_{0,11} )$ is the transition
frequency between the two energy levels and
$\Omega_\mathrm{R} = \hbar^{-1} \mu_{z,21} E_z$ is the instantaneous Rabi
frequency.
Furthermore, the dephasing rate
\begin{equation}
  \gamma_2 = \frac{1}{2}
  \left( \gamma_{12} + \gamma_{21} \right) + \gamma_{12,\mathrm{p}},
\end{equation}
the scattering rate $\gamma_1 = \gamma_{12} + \gamma_{21}$, and the
equilibrium population inversion
\begin{equation}
  w_0 = \frac{\gamma_{21} - \gamma_{12}}{\gamma_{21} + \gamma_{12}}
\end{equation}
are introduced to simplify the terms induced by the Lindblad superoperator.

In order to provide a convenient alternative for the Maxwell-Bloch equations
in that form to the user, we implemented a subclass \clsn{qm\_desc\_2lvl}
whose constructor accepts six real values.
Those values represent the density $n_{3D}$, the transition frequency, the
dipole length $z_{21} = -e^{-1} \mu_{z,21}$ (where $e$ is the elementary charge),
the scattering rate, the dephasing rate, and the equilibrium inversion value,
respectively.
Then, the constructor builds the Hamiltonian
\begin{equation}
  \hat H_0 = \frac{\hbar \omega_{21}}{2}
  \begin{bmatrix} -1 & 0\\ 0 & 1 \end{bmatrix},
\end{equation}
the dipole operator
\begin{equation}
  \hat \mu_z = -e z_{21}
  \begin{bmatrix} 0 & 1\\ 1 & 0 \end{bmatrix},
\end{equation}
and the Lindblad superoperator.
For the latter step, the scattering rate matrix
\begin{equation}
  R =
  \begin{bmatrix}
    0 & \gamma_{12}\\
    \gamma_{21} & 0
  \end{bmatrix} =
  \frac{\gamma_1}{2}
  \begin{bmatrix}
    0 & 1 - w_0\\ 1 + w_0 & 0
  \end{bmatrix}
\end{equation}
and the pure dephasing rate $\gamma_{12,\mathrm{p}} = \gamma_2 - \gamma_1/2$
have to be determined.

At this point, we have described the regions and materials of the device. Now
we need to include the boundary conditions. As mentioned above, it is
sufficient to store two real values that represent the reflectivity values of
both ends of the device. Those values could be integrated directly into the
device class. However, in order to maintain the flexible nature of our base
library, we decided to add two pointers to an abstract class \clsn{bc\_field}
to the device class. Thereby, the project can be extended easily in future,
e.g., to incorporate periodic boundary conditions. At the moment, the
only subclass of the abstract class is \clsn{bc\_field\_reflectivity}, whose
constructor accepts a real reflectivity value.

\subsubsection{Scenario setup and initial conditions}

As outlined above, the class \clsn{scenario} contains the dynamical part
of the simulation setup. Namely, those are the source terms and the initial
conditions. Although in most examples one source term is sufficient, the
\clsn{scenario} can contain any number of terms, which are stored as
pointers to the class \clsn{source}. Similar to other classes mentioned
before, \clsn{source} is a base class that stores common information, such
as the position $x$ at which the source should be placed. Also, the
\clsn{source} features a type field that distinguishes hard and soft
sources. Here, it should be noted that a hard source sets the value of the
electric field to the source value, whereas the soft source adds the source
value to the current field value~\cite{taflove2005}. Different subclasses
can be derived from the class \clsn{source}, such as \clsn{sech\_pulse}
and \clsn{gaussian\_pulse}. As the name suggests, those subclasses
yield a sech and a Gaussian pulse, respectively.

Similar to the treatment of the boundary conditions in the device, the
scenario contains pointers to abstract classes that represent the initial
conditions. Here, the pointer to \clsn{ic\_density} specifies the
initialization of the density matrix, and two pointers to \clsn{ic\_field}
determine the initial values of electric and magnetic field, respectively.
Currently, only a subclass \clsn{ic\_density\_const}, which yields a
constant initial density matrix, is implemented. For the fields there are
more options: constant initialization, random initialization, and even a
certain initial curve can be specified.

Apart from the simulation setup, further properties can be specified. For
example, the number of spatial grid points can be specified. Thereby, the
user can increase the accuracy and determine the effect on the results
itself, as well as on the performance of the solver. Finally, the user
needs to specify the desired results. Even in the most trivial simulations,
several data sets are generated that are not required. In order to avoid
wasting memory, we added a collection of \clsn{record} objects to the
scenario. Each record specifies a certain quantity that should be recorded,
and contains information on the sampling interval and position. Then, during
the simulation run, the solver analyses the information in the record list
and stores the corresponding data traces in \clsn{result} objects. The latter
are data container classes, which can be analyzed during postprocessing
(either by accessing them in system memory or after exporting them to a
file).

\subsubsection{Solver and writer infrastructure}

The base library only contains the abstract classes \clsn{solver} and
\clsn{writer}, and leaves the implementation to libraries that build on the
base. Before we discuss the resulting plugin structure in more detail, let
us take a look at the common properties of all solvers and writers,
respectively, that are represented by the abstract classes.
The constructor of \clsn{solver} expects the name of the solver, as well as
the device and the scenario to be simulated. After the solver is created,
the method \clsn{run} executes the simulation. Then, the results can be
extracted with the method \clsn{get\_results}. The abstract class
\clsn{writer} features a method \clsn{write} that accepts the results and
writes them to a file. In addition to the results, the target filename, the
device, and the scenario must be specified. The latter are required since
the simulation result files should contain meta-information, such as the name
of the device and the discretization size.

The plugin structure mentioned above guarantees the required flexibility.
For the sake of brevity, we describe this approach for the solver. All
remarks in the following hold analogously for the writer. The constructor
of \clsn{solver}, which we already introduced, is indeed marked as protected.
This means that instances of this class cannot be created directly. In order
to create an instance of a certain solver, the static method
\clsn{create\_instance} must be called. This method expects the name of the
solver as parameter (in addition to device and scenario), looks up the
corresponding subclass of \clsn{solver}, and returns an instance of this
subclass (using the provided device and scenario). While this approach may
seem overly complex at first glance, it provides a clean interface to the
user. For example, the user can acquire the available solvers with the
static method \clsn{get\_avail\_solvers} and choose to create an instance
of one of them without knowing the name of the corresponding subclass.
In the following, we present one concrete implementation of \clsn{solver}
and \clsn{writer}, respectively.

\subsection{solver-cpu: Solver based on the OpenMP standard}

Currently, mbsolve contains one solver implementation, which targets CPU
based shared memory systems. The key elements of this implementation -- in
the following referred to as \clsn{solver-cpu} -- are outlined briefly to
give the user some initial guidance.

At the creation of a \clsn{solver-cpu}
object, the number of spatial and temporal grid points is determined.
Typically, the user specifies a certain number of spatial grid points $N_x$.
Then, the discretization size $\Delta x = L/(N_x - 1)$ can be calculated for
a given total device length $L$. It should be noted that the user must take
care of the spatial discretization size, which should be chosen in the range
$\lambda/20$ to $\lambda/200$, where $\lambda$ is the smallest occurring
wavelength~\cite{jirauschek2019ats}. As next step, the temporal
discretization size $\Delta t = C \Delta x/c$ is calculated, where $c$ is the
speed of light, and $C < 1$ is the Courant number. Since the speed of light may
vary between the different materials, we select the largest occurring value
for $c$, which results in a minimal $\Delta t$. At the moment, the Courant
number cannot be chosen by the user. It is set to $C = 1/2$, which was found
to be adequate in related literature~\cite{jirauschek2019ats}.
Based on the simulation end time $t_\mathrm{e}$ the number of temporal grid
points can be calculated, where the discretization size $\Delta t$ may be
decreased slightly to allow an integer number $N_t$.
We note that for the borderline case $N_x = 1$ the spatial discretization
size calculation does not make sense. In this case, the user can specify the
number of temporal grid points, which determines the temporal discretization
size.
For convenience reasons, the complete process is delegated to the helper
function \clsn{init\_fdtd\_simulation}, which analyses the given device and
scenario and calculates appropriate numbers of grid points.

Subsequently, the material parameters are converted to coefficients in
order to avoid unnecessary, yet costly operations in the following
simulation run. After rearranging Eqs.~(\ref{eq:ediscrete}) and
(\ref{eq:hdiscrete}), we find that the update equations now read
\begin{equation}
  \begin{aligned}
    E_z^{m, n+1} =& a' E_z^{m, n}
    - b' \Gamma \left( \partial_t P_z\right)^{m, n + 1/2} +\\
    & b' \Delta x_{\mathrm{inv}}
    \left( H_y^{m + 1/2,n + 1/2} - H_y^{m - 1/2,n + 1/2} \right),
  \end{aligned}
  \label{eq:eupdate}
\end{equation}
where the coefficients
\begin{equation*}
  a' = \frac{1 - \Delta t (2\epsilon)^{-1}\sigma}{1 +
    \Delta t (2\epsilon)^{-1}\sigma}, \qquad
  b' = \frac{\Delta t\epsilon^{-1}}{1 + \Delta t (2\epsilon)^{-1}\sigma}, \qquad
  \Delta x_{\mathrm{inv}} = \frac{1}{\Delta x},
\end{equation*}
are used, and
\begin{equation}
  H_y^{m + 1/2,n + 1/2} =
    H_y^{m + 1/2,n-1/2} + c' \left( E_z^{m + 1, n} - E_z^{m, n} \right),
  \label{eq:hupdate}
\end{equation}
where the coefficient
\begin{equation*}
  c' = \frac{\Delta t}{\Delta x \mu}.
\end{equation*}
We note that the coefficient $\Delta x_{\mathrm{inv}}$ is constant in the
whole structure. It is therefore precalculated once to save costly division
operations. The coefficients $a'$, $b'$, and $c'$ are constant in each
material.
They are precalculated by the helper function \clsn{get\_fdtd\_constants}
and stored in simple data structures.

As next step, the data structures that store intermediate values are
created. Namely, those intermediate values are the electric field, the
magnetic field, the polarization term $\partial_t P_z$, and the density
matrix. Those values are stored for a single time step, but for the complete
spatial grid. After creation, the data structures are initialized according
to the given initial conditions.
Care must be taken that those structures are cleaned up appropriately in the
destructor, i.e., when the solver object is destroyed.

Finally, the specified record and source objects are converted into data
structures that are more convenient. Most importantly, the specified position
and time interval values are converted into integer numbers, which express
the values in terms of the discrete spatiotemporal grid.

At this point, the solver is successfully created and its method \clsn{run}
can be called. This method executes the simulation loop, which is outlined
in Algorithm~\ref{alg:loop_basic}. The simulation loop iterates over the
temporal grid points and updates the intermediate values of electric field
$e$, magnetic field $h$, density matrix $d$, and polarization term $p$ from
time step to time step.
In each iteration, there are two nested loops, which iterate over
the spatial grid points.
The first nested loop updates the quantities that are discretized at $n+1/2$
(namely, those are the magnetic field, the density matrix, and the
polarization term).
The second updates the electric field, which is discretized at $n$. Due to
the data dependencies between the nested loops, synchronization mechanisms
are required that guarantee that the first loop is completed before the
second loop starts.
Within the nested loops, however, there are no data dependencies, which
allows them to be executed in parallel. With the help of the OpenMP standard
it was quite straightforward to achieve this. Unfortunately, this
straightforward solution has a drawback. In cases where the computational
workload per spatial grid point is little, the calls to the synchronization
routine result in increasing idle time, which of course impedes the
parallel efficiency. For those cases, an advanced approach using redundant
calculations is beneficial. Both approaches are described in more detail
in~\cite{riesch2019pasc}.

\begin{algorithm}
\centering
\begin{algorithmic}
  \FOR{$n = 0$ \TO $n_\mathrm{max}$}

  \FOR{$m = 1$ \TO $m_\mathrm{max}$}
  \STATE{$h[m] \gets \mathrm{update\_h}(e[m], e[m-1])$}
  \STATE{$d[m] \gets \mathrm{update\_d}(e[m])$}
  \STATE{$p[m] \gets \mathrm{calc\_p}(d[m])$}
  \ENDFOR
  \STATE{sync()}

  \FOR{$m = 0$ \TO $m_\mathrm{max}-1$}
  \STATE{$e[m] \gets \mathrm{update\_e}(h[m + 1], h[m], p[m])$}
  \ENDFOR
  \STATE{sync()}
  \STATE{record\_results()}
  \ENDFOR
\end{algorithmic}
\caption{Simulation main loop -- basic version~\cite{riesch2019pasc}.}
\label{alg:loop_basic}
\end{algorithm}

We recall that different numerical methods (and variations thereof) are
likely to be implemented in mbsolve.
Since it is considered bad practice to copy and paste parts of the code,
we need a way to factor out the common parts.
The helper functions mentioned above are one example for such common code.
Other examples are the routines to update the density matrix and to
calculate the polarization contribution, as they are the same for both the
basic as well as the advanced version of the simulation loop.
Conversely, one version of the simulation loop could be tested with different
variants of the density matrix update step.
Therefore, we created two classes for the currently available density matrix
algorithms, namely \clsn{lindblad\_reg\_cayley} for the operator-splitting
approach from~\cite{bidegaray2001} and \clsn{lindblad\_cvr\_rodr} for the
numerical method presented in~\cite{riesch2017c}.
Those classes could be connected to the different solver variants (e.g.,
\clsn{solver\_cpu\_fdtd} for the basic simulation loop, and
\clsn{solver\_cpu\_fdtd\_red} for the advanced version) at compile time or
runtime, where the former approach is more complex.
However, it opens the opportunity for compile time optimization,
which is the reason why we decided to introduce template class arguments
for the solver variants.
As a consequence, different template instances can be created at compile
time by specifying the number of energy levels and the algorithm class.

Finally, at the end of each temporal iteration the record data structures
are analyzed.
In Algorithm~\ref{alg:loop_basic}, this procedure is denoted
\clsn{record\_results} for simplicity, although it is not an actual
function.
In this part of the code, a decision is made if and which intermediate
results are recorded, and, where appropriate, the data are written in
the result objects.

\subsection{writer-hdf5: Writer for the Hierarchical Data Format (HDF5)}

As already mentioned, the result objects can be passed to a writer.
Currently, the only available writer implementation in mbsolve stores the
data in the HDF5 format. This format is well accepted in computational
science, and is supported by most programming languages (including C++,
Python, MATLAB, and Octave) on all major platforms. The three main entities
of the HDF5 format are groups, data sets, and attributes. Using groups,
a hierarchical structure can be created. In each group (including the root),
data sets and attributes can be placed, where attributes can be used to
store meta-information.

The \clsn{writer-hdf5} stores the simulation meta-information (e.g., temporal
discretization size) in attributes of the root group. Then, it creates a
separate group for each result. Naturally, a data set containing
the result data is added to this newly created group. However, since HDF5
does not natively support complex numbers, a second data set has to be added
in case the result is complex (e.g., off-diagonal entries of the density
matrix). Additionally, the writer creates a per-result attribute that
informs the user whether the result is complex or real.

\subsection{Project management and software quality assurance}

From the very beginning of the mbsolve project, we strived to provide a
reliable solution of high software quality to the scientific community.
Indeed, any scientific software package must work reliably, as it
serves as third pillar of science, the other two being theory and experiment.
Clearly, creating a reliable solution requires significant effort, which is
also one reason why we chose to make the resulting source code publicly
available, in the hope that our efforts will help other research groups
as well.

During the development of the software presented in this paper, we
identified best practices in (scientific) software development and
implemented them. In order to provide guidance for other software projects,
we compiled our findings into a project skeleton~\cite{riesch2020pone}. This
skeleton can be used to create new software project with a few button clicks.
The newly created project features the implementations of several best
practices from the very start. In the following, we describe briefly the
best practices, which are also used in the mbsolve project.

One elementary best practice is using a version control system. While it is
beneficial even for a single user scenario, it becomes indispensable as soon
as more developers work on the project. Ideally, it is combined with a
collaboration web tool, which also offers an issue tracking system. We
decided to host the source code on GitHub, which offers both.
As to the quality of the source code, we defined a coding convention for our
project and established checks whether new contributions are compliant with
it. During the design of the software, we aimed to provide a clear and
object-oriented architecture. We exploited modern C++ features, such as
smart pointers that prevent memory leaks, in order to avoid typical
mistakes, and to produce readable code. As far as third party components
are concerned, we selected open-source components exclusively.
Additionally, the source code can be compiled with a variety of compilers
(open-source and proprietary) on the three major operating systems Linux,
Windows, and macOS.
By using continuous integration (CI), we automated repetitive tasks such as
building the project, testing the resulting libraries, and performing
additional tasks.
For example, the documentation is generated based on comments in the source
code, compiled to static HTML pages, and uploaded to GitHub
Pages~\cite{mbsolve-doc}.

\subsection{Installation and requirements}

As we want to invite other researchers to use and extend the mbsolve
software, we describe in the following the build tools, the third party
dependencies, and the steps required to build and/or install the software.

In case the mbsolve software should be built from the source code, the CMake
build system (version $\geq$ 3.9) and a recent C++ compiler is required. As
to the latter, we have compiled the mbsolve source code successfully using
the GNU Compiler Collection (gcc, version $\geq$ 4.9.2), the Intel C++
Compiler (any recent version), the Clang compiler (version $\geq$ v7.0.0),
and the Microsoft Visual C++ compiler (MSVC, any recent version). It should
be noted that AppleClang lacks OpenMP support entirely, and MSVC only
supports a dated version, which may result in inferior performance. Then,
the Eigen library (version $\geq$ 3.3.4) and the HDF5 libraries (any recent
version) must be installed for the \clsn{solver-cpu} and \clsn{writer-hdf5},
respectively. The Python interface is optional, and requires Python (version
$\geq$ 2.7) and SWIG (version $\geq$ 2.0.12). Finally, cxxopts (any recent
version) is a prerequisite for the optional mbsolve-tool, and Doxygen (any
recent version) is required for generating the documentation.
If the required dependencies are not met, certain components may be disabled
automatically by the build system.

Once the requirements are set up, the build process consists of executing
CMake, which creates the project files for a certain generator, and running
the generator, which depends on the platform in use (GNU make, Microsoft
Visual Studio, etc.). Please refer to the documentation~\cite{mbsolve-doc}
for a more detailed description.

Alternatively, the compiled binaries can be installed directly. Here, the
dependencies are a reduced set of the list above. Namely, those are the
C/C++ standard libraries including OpenMP support, the Python runtime, and
the HDF5 libraries. However, the same versions as used during compilation
must be available, which is far from trivial. Therefore, we recommend the
installation of the binaries via conda, where the command
\begin{verbatim}
  $ conda install -c conda-forge mbsolve
\end{verbatim}
installs the mbsolve binaries together with all required dependencies. This
approach works for all major platforms (Linux, Windows, and macOS), although
the way to install a conda package may differ.

\section{Applications and simulation examples}
\label{sec:applications}

In the following, we demonstrate the usage of the mbsolve project with the
help of four application examples. Those examples have been selected so that
they represent different simulation types (the Maxwell-Bloch equations in 1D,
or solving only the master equation at a certain point in space),
use different features of mbsolve (different source types, initial conditions,
etc.),
and are executed using the C++ interface as well as the Python bindings.
Furthermore, the selected simulations feature different numbers of energy
levels, and while three examples are rather of theoretical nature, the final
simulation models a real quantum cascade laser (QCL).
Apart from providing initial guidance to the prospective user, the examples
serve as verification of the implementation.

\subsection{Self-induced transparency (SIT) in two-level systems}

\begin{figure}
  \centering
  \includegraphics[width=0.8\linewidth]{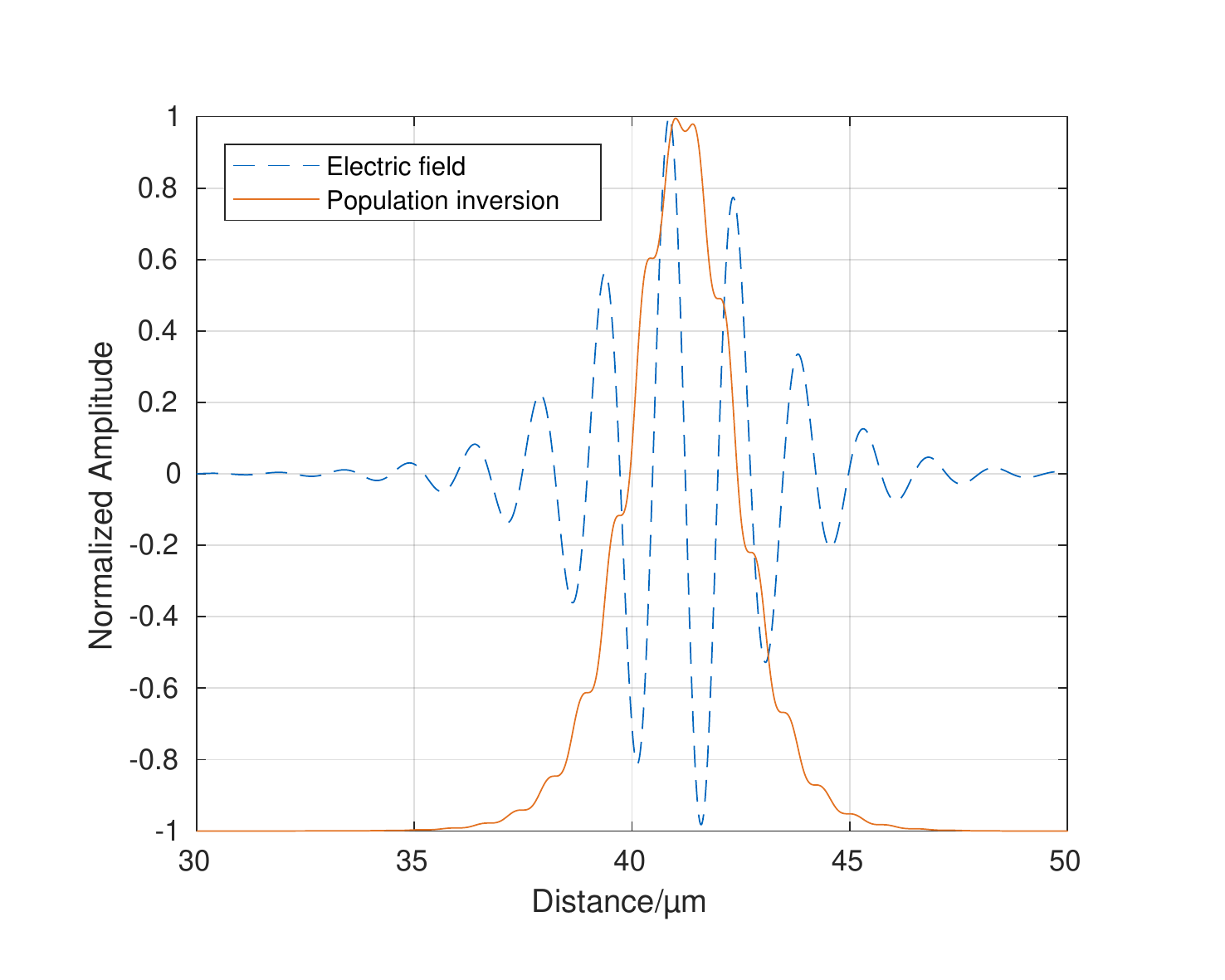}
  \caption{Simulation results from the self-induced transparency simulation
    setup, cf.~Ziolkowski et al.~\cite{ziolkowski1995}, Fig. 2.}
  \label{fig:result_ziolkowski}
\end{figure}

The first application example reproduces the results presented in the work
by Ziolkowski et al.~\cite{ziolkowski1995}, who (to the best of our knowledge)
were the first to use the FDTD method for Maxwell-Bloch equations.
In this pioneering work, the self-induced transparency (SIT) effect in
two-level systems is investigated.
Here, the active region is embedded in two short vacuum regions.
In one vacuum region a sech pulse is injected and subsequently travels
through the active region.
By setting the pulse area to $\pi$, $2\pi$, and $4\pi$, the quantum
mechanical systems are inverted once, twice, and four times, respectively.
Figure~\ref{fig:result_ziolkowski} depicts the population inversion and the
electric field for a $2\pi$ pulse, after it has propagated for
\SI{200}{\femto\second}.

Listing~\ref{lst:ziolkowski} shows the complete Python code required to set
up and run the SIT example.
After the mbsolve libraries are imported, the script creates two materials,
one of which features a simple two-level quantum mechanical description.
The device to be simulated contains three regions, where vacuum is assigned
to the first and third regions, and the active region is put in the middle.
Initially, all quantum mechanical systems are inverted, which is represented
by the initial density matrix \texttt{rho\_init}.
The scenario specifies 32768 spatial grid points, and that the simulation is
run for \SI{200}{\femto\second}.
Furthermore, two records are added that request the recording of the
population inversion and the electric field, respectively.
Both quantities are to be sampled using a time interval of
\SI{2.5}{\femto\second} over the complete spatial domain.
Also, a source, which represents a sech pulse at the left end of the device,
is added to the scenario.
Finally, the solver is created and runs the simulation.
On a recent quad-core desktop computer this simulation should be processed
in less than 20 seconds.
The results are passed to the writer, which stores the data in a HDF5 file.

\pagebreak

\begin{lstlisting}[language=Python,caption={Python script for the
  self-induced transparency (SIT) setup,
  cf.~\texttt{tools/python/ziolkowski1995.py} in the mbsolve
  repository~\cite{mbsolve-github}.},label={lst:ziolkowski}]
# import mbsolve libraries
import mbsolve.lib as mb
import mbsolve.solvercpu
import mbsolve.writerhdf5

# vacuum
mat_vac = mb.material("Vacuum")
mb.material.add_to_library(mat_vac)

# simple two-level quantum mechanical description
# params: particle density, transition frequency,
#         transition dipole length,
#         scattering rate gamma_1, dephasing rate gamma_2,
#         equilibrium population inversion w_0
qm = mb.qm_desc_2lvl(1e24, 2 * math.pi * 2e14, 6.24e-11, 1.0e10, 1.0e10,
                     -1.0)

# Ziolkowski active region material
mat_ar = mb.material("AR_Ziolkowski", qm)
mb.material.add_to_library(mat_ar)

# Ziolkowski device setup
dev = mb.device("Ziolkowski")
# params of each region: name, material, x_start, x_end
dev.add_region(mb.region("Vacuum left", mat_vac, 0, 7.5e-6))
dev.add_region(mb.region("Active region", mat_ar, 7.5e-6, 142.5e-6))
dev.add_region(mb.region("Vacuum right", mat_vac, 142.5e-6, 150e-6))

# initial density matrix
# params: lower level rho_11 fully populated, upper level rho_22 empty
#         vector with coherence terms not set in params, zero by default
rho_init = mb.qm_operator([ 1, 0 ])

# scenario
ic_d = mb.ic_density_const(rho_init)
ic_e = mb.ic_field_const(0.0)
sce = mb.scenario("Basic", 32768, 200e-15, ic_d, ic_e)
# record electric field and population inversion in 2.5fs intervals
sce.add_record(mb.record("inv12", 2.5e-15))
sce.add_record(mb.record("e", 2.5e-15))

# add source
sce.add_source(mb.sech_pulse("sech", 0.0, mb.source.hard_source,
                             4.2186e9, 2e14, 10, 2e14))

# run solver (advanced FDTD implementation + approach by Bidegaray 2001)
sol = mb.solver.create_instance("cpu-fdtd-red-2lvl-reg-cayley", dev,
                                sce)
sol.run()

# write results
wri = mb.writer.create_instance("hdf5")
outfile = dev.get_name() + "_" + sce.get_name() + "." +
          wri.get_extension()
results = sol.get_results()
wri.write(outfile, sol.get_results(), dev, sce)
\end{lstlisting}

\subsection{Pulse propagation in a V-type three-level system}

Using Maxwell-Bloch simulations, Song et al.~\cite{song2005} investigated
a setup that is conceptually similar to the SIT example, the major
difference being the active region.
Here, atomic rubidium was modeled as a three-level quantum mechanical system.
While the work mainly focused on the propagation of few-cycle pulses, it also
contains the temporal simulation of a single three-level system that is
driven by a pulse (cf.~\cite{song2005}, Fig.~3).
This case can be reproduced by solving the Lindblad master equation alone.

In Listing~\ref{lst:song}, the creation of a device and a scenario for this
setup is given.
Since a three-level system is considered here, we have to use the complete
quantum mechanical description including the operators and superoperator.
In this example, the device contains only one region of zero length.
This is the first indication that the simulation does not consider any
spatial dimensions.
The second indication is the number of grid points, which is passed to the
scenario constructor.
Here, the number of spatial grid points is set to 1, whereas the 10000
temporal grid points are specified.
Similar to the SIT setup, records and sources can be added, although it only
makes sense to add them at the position of the single grid point.
It should be noted that for a low number of spatial grid points
parallelization of the calculation does not make sense.
Fortunately, it is not needed here as this example will complete within
seconds on a recent desktop computer (using only one core).

\begin{lstlisting}[language=Python,caption={Code snippet of the Python script
  that reproduces the three-level driven quantum mechanical system in
  \cite{song2005},
  cf.~\texttt{tools/python/song2005.py} in the mbsolve
  repository~\cite{mbsolve-github}.},label={lst:song}]
# Hamiltonian
# (diagonal qm operator, vector for off-diagonal terms omitted)
energies = [ 0, 2.3717e15 * mb.HBAR, 2.4165e15 * mb.HBAR ]
H = mb.qm_operator(energies)

# dipole moment operator (qm operator with off-diagonal terms)
dipoles = [ -mb.E0 * 9.2374e-11, -mb.E0 * 9.2374e-11 * math.sqrt(2), 0]
u = mb.qm_operator([ 0, 0, 0 ], dipoles)

# relaxation superoperator
rate = 1e10
# scattering rate matrix R
rates = [ [ 0, rate, rate ], [ rate, 0, rate ], [ rate, rate, 0 ] ]
# pure dephasing rates are zero in this example
pure_deph = [ 0, 0, 0 ]
relax_sop = mb.qm_lindblad_relaxation(rates, pure_deph)

# initial density matrix
rho_init = mb.qm_operator([ 1, 0, 0 ])

# quantum mechanical description
qm = mb.qm_description(6e24, H, u, relax_sop)
mat_ar = mb.material("AR_Song", qm)
mb.material.add_to_library(mat_ar)

# Song setup
dev = mb.device("Song")
dev.add_region(mb.region("Active region (single point)", mat_ar, 0, 0))

# scenario
ic_d = mb.ic_density_const(rho_init)
ic_e = mb.ic_field_const(0.0)
ic_m = mb.ic_field_const(0.0)
sce = mb.scenario("Basic", 1, 80e-15, ic_d, ic_e, ic_m, 10000)
sce.add_record(mb.record("e", 0.0, 0.0))
sce.add_record(mb.record("d11", mb.record.density, 1, 1, 0.0, 0.0))
sce.add_record(mb.record("d22", mb.record.density, 2, 2, 0.0, 0.0))
sce.add_record(mb.record("d33", mb.record.density, 3, 3, 0.0, 0.0))

# add source
sce.add_source(mb.sech_pulse("sech", 0.0, mb.source.hard_source,
                             3.5471e9, 3.8118e14, 17.248, 1.76/5e-15,
                             -math.pi/2))
\end{lstlisting}

\subsection{Six-level anharmonic ladder system}

In the work by Marskar and \"Osterberg~\cite{marskar2011}, two simulation
examples are discussed.
The first example is a variation of the SIT setup in~\cite{ziolkowski1995},
which we have already discussed above.
The second example considers the propagation of a Gaussian pulse in a medium
that is modeled as a six-level anharmonic ladder system
(cf.~\cite{marskar2011}, Fig.~4).
In this system, the energy levels are given as
\begin{equation}
  E_{n + 1} = E_{n} + \hbar \omega_0 [ 1 - 0.1 (n - 3) ],
\end{equation}
where $n \in [1;N-1]$, $N$ is the number of energy levels, and
$\omega_0 = \SI{2 \pi e13}{\per\second}$ is the transition frequency between
the energy levels $E_3$ and $E_4$.
For convenience, and without loss of generality, we set $E_1 = 0$ to zero.

From our perspective, this setup is no more than an variation of the previous
examples.
However, it gives us a nice opportunity to introduce the mbsolve-tool, which
features different simulation examples written in C++.
In fact, all examples discussed in this section can be started by specifying
the setup name as well as an appropriate solver and writer as command line
arguments.
For example, the anharmonic ladder simulation can be started as
\begin{verbatim}
  $ mbsolve-tool -w hdf5 -m cpu-fdtd-red-6lvl-reg-cayley -d marskar2011-6lvl
\end{verbatim}
Depending on the given setup name, the application creates the corresponding
device and scenario, runs the specified solver, and uses the given writer
to store the simulation results.
Further command line arguments can be used to specify the number of spatial
grid points and the simulation end time.
This feature was extensively used during performance
tests~\cite{riesch2018oqel}.
By default, 8192 spatial grid points and a simulation end time of
\SI{2}{\nano\second} are used, resulting in a runtime of approximately
two minutes on a recent quad-core desktop computer.

We note that in the command line entry above, the number of energy levels
seems to be stated explicitly as part of the device name.
Indeed, the name ``marskar2011-6lvl'' and the shortcut ``marskar2011'' refer
to the original setup, but in fact any number $N \geq 2$ can be specified.
This is a feature we added for performance comparisons of numerical methods,
in which the performance is analyzed with respect to the number of energy
levels (e.g., in~\cite{riesch2019cleo1,riesch2020nusod1}).
While the generalized example may not necessarily make sense from the
modeling point of view, it represents a typical application example and hence
constitutes a reasonable benchmark.

\subsection{Quantum cascade laser frequency comb}

Finally, we present the most complex and computationally demanding
application example, as one simulation run required up to two hours on
an AMD Ryzen Threadripper 2990WX machine using 16 cores.
The quantum cascade laser frequency comb presented in \cite{burghoff2014} was
modeled in previous work of our group~\cite{tzenov2016}, where the
experimental results were reproduced with good agreement.
Most input parameters were determined using prerequisite Schrödinger-Poisson
and ensemble Monte Carlo simulations.
Thereby, the number of empirical model parameters were reduced to a minimum.

Listing~\ref{lst:qcl} shows how a similar simulation can be set up using
the mbsolve software.
First, the quantum mechanical description of the active region material is
created.
Apart from the five eigenenergies on the main diagonal, the Hamiltonian in
this example has non-zero off-diagonal elements, which account for tunneling
effects.
All the elements are determined by a Schrödinger-Poisson simulation in
tight-binding basis~\cite{jirauschek2014}.
The dipole moment operator is less spectacular as it only contains one
non-zero element, which is placed on the off-diagonal element that corresponds
to the transition between upper and lower laser level.
The scattering rates and dephasing rates are used to set up the Lindblad
relaxation superoperator.
The quantum mechanical description and the active region material are then
created using the parameters from~\cite{tzenov2016}.
Then, semi-transparent mirror boundary conditions are created, with the
reflectivity values $R_1 = R_2 = 0.8$.
The boundary conditions are subsequently used during the creation of the
device, to which one region of the active region material is added.
After that, the scenario is set up using the initial density matrix, in
which only the upper laser level is populated.
The electric field is initialized randomly to model stimulated emission,
whereas the magnetic field is set to zero.
Since this is the default in mbsolve, we do not need to specify this choice
of initial conditions explicitly.
Finally, a record that triggers the recording of the electrical field at the
facet is added to the scenario.

\begin{lstlisting}[language=C++,caption={Code snippet of the mbsolve-tool
  C++ application that reproduces the quantum cascade laser frequency comb
  simulation in~\cite{tzenov2016},
  cf.~\texttt{mbsolve-tool/src/mbsolve-tool.cpp} in the mbsolve
  repository~\cite{mbsolve-github}.},label={lst:qcl}]
/* quantum mechanical description of active region */
/* params: vector of main diagonal entries, followed by vector of
 * off-diagonal entries */
mbsolve::qm_operator H(
  { 0.10103 * mbsolve::E0, 0.09677 * mbsolve::E0, 0.09720 * mbsolve::E0,
    0.08129 * mbsolve::E0, 0.07633 * mbsolve::E0 },
  { 0.0, 1.2329e-3 * mbsolve::E0, -1.3447e-3 * mbsolve::E0,
    0.0, 0.0, 0.0, 0.0, 0.0, 0.0, 0.0 });

mbsolve::qm_operator u(
  { 0.0, 0.0, 0.0, 0.0, 0.0 },
  { 0.0, 0.0, 0.0, 0.0, 0.0, -mbsolve::E0 * 4e-9, 0.0, 0.0, 0.0, 0.0 });

/* the scattering rate matrix R */
std::vector<std::vector<mbsolve::real> > scattering_rates = {
  { 0.0000000, 0.4947e12, 0.0974e12, 0.8116e12, 1.0410e12 },
  { 0.8245e12, 0.0000000, 0.1358e12, 0.6621e12, 1.1240e12 },
  { 0.0229e12, 0.0469e12, 0.0000000, 0.0794e12, 0.0357e12 },
  { 0.0047e12, 0.0029e12, 0.1252e12, 0.0000000, 0.2810e12 },
  { 0.0049e12, 0.0049e12, 0.1101e12, 0.4949e12, 0.0000000 }
};

mbsolve::real deph_inj_ull = 1.0/(0.6e-12);
mbsolve::real deph_xxx_xxx = 1.0/(1.0e-12);

/* the vector of pure dephasing rates */
std::vector<mbsolve::real> dephasing_rates =
  { 0, deph_inj_ull, deph_inj_ull, deph_xxx_xxx, deph_xxx_xxx,
    deph_xxx_xxx, deph_xxx_xxx, deph_xxx_xxx, deph_xxx_xxx,
    deph_xxx_xxx };

auto relax_sop = std::make_shared<mbsolve::qm_lindblad_relaxation>(
    scattering_rates,
    dephasing_rates);

auto qm = std::make_shared<mbsolve::qm_description>(5.6e21, H, u,
                                                   relax_sop);

auto mat_ar = std::make_shared<mbsolve::material>("AR", qm, 12.96, 0.9,
                                                  1100);

/* set up device with semi-transparent mirror boundary cond. */
auto bc =
    std::make_shared<mbsolve::bc_field_reflectivity>(0.8, 0.8);
dev = std::make_shared<mbsolve::device>("tzenov2016", bc);
dev->add_region(std::make_shared<mbsolve::region>(
"Active region", mat_ar, 0, 5e-3));

/* initial density matrix (rho_33 is fully populated) */
mbsolve::qm_operator rho_init({ 0.0, 0.0, 1.0, 0.0, 0.0 });

/* basic scenario */
scen = std::make_shared<mbsolve::scenario>(
    "basic", num_gridpoints, sim_endtime, rho_init);
scen->add_record(std::make_shared<mbsolve::record>(
    "e0", mbsolve::record::electric, 1, 1, 0.0, 0.0));
\end{lstlisting}

After a solver has processed the device and scenario, the result data are
written to a HDF5 file.
At this point, we would like to discuss briefly the postprocessing of results.
Typically, several tasks remain to be done after the solver has completed.
In this case, the Fourier transform of the recorded electric field must be
calculated and plotted.
Naturally, this is beyond the scope of mbsolve and established software
tools, such as MATLAB, Octave, or Python (with NumPy, SciPy, and Matplotlib),
should be used.
Examples for postprocessing MATLAB scripts can be found in
\texttt{tools/matlab} in the mbsolve repository~\cite{mbsolve-github}.

\begin{figure}
  \centering
  \includegraphics[width=0.8\linewidth]{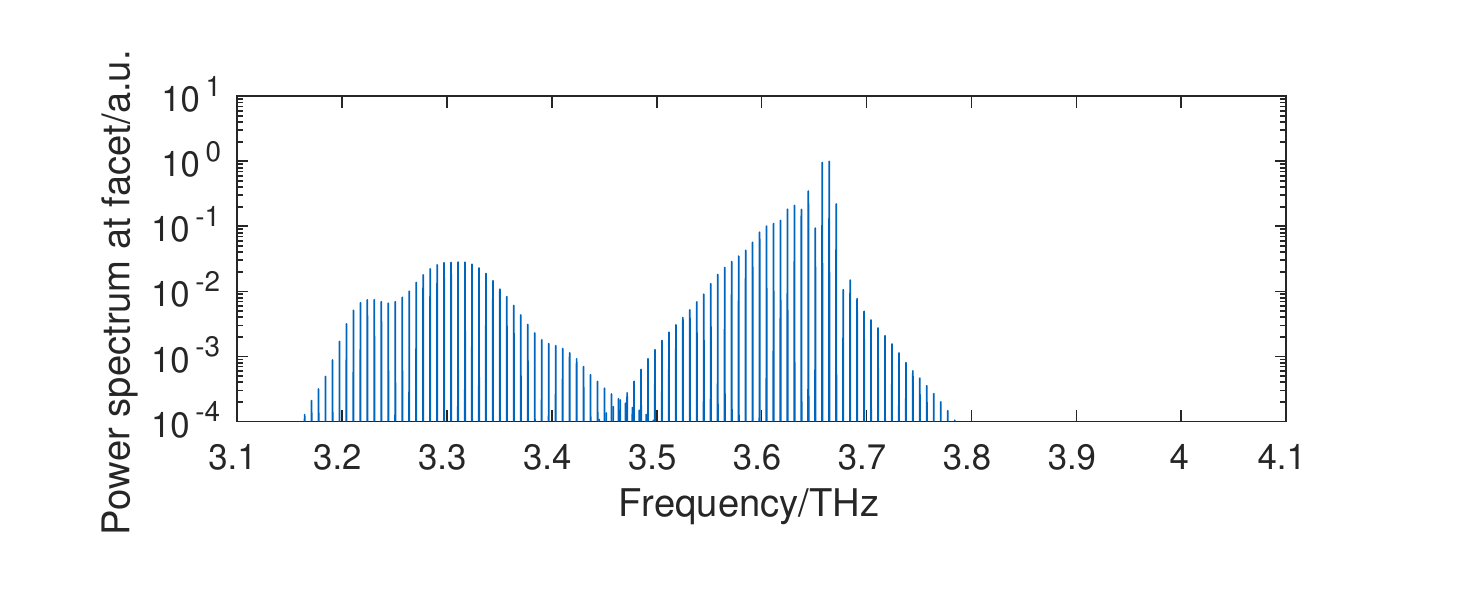}
  \caption{Power spectrum of the electric field recorded at the facet of the
    quantum cascade laser frequency comb.
    This simulation example bases on Tzenov et al.~\cite{tzenov2016}
    (cf.~Fig.~5c).
    For the corresponding experimental results cf.~Burghoff et
    al.~\cite{burghoff2014}, Fig.~3b.}
  \label{fig:result_qcl}
\end{figure}

The simulation results (after postprocessing) are depicted in
Fig.~\ref{fig:result_qcl}.
While the spectrum of the electric field at the facet shows reasonable
agreement with the experiment, there are features missing that the simulation
in~\cite{tzenov2016} could capture.
We believe that this is due to the differences in the underlying simulation
methods.
While the method in this work does not invoke the RWA, it does not
account for chromatic dispersion or diffusion processes due to spatial hole
burning.

\section{Conclusion}
\label{sec:conclusion}

In this paper, we have presented mbsolve, an open-source solver tool for
the Maxwell-Bloch equations.
Here, we have focused on the one-dimensional case, which is suitable for
e.g., modeling various types of optoelectronic devices such as QCLs, where
the waveguide geometry allows a reduction of the model to one spatial
dimension.
Besides optoelectronic devices, the resulting equations can be applied to
a variety of problems in which the plane-wave approximation is a
reasonable assumption.
Also, we have outlined the numerical treatment of the Maxwell-Bloch equations,
focusing on methods that do not invoke the RWA/SVEA.
Additionally, the methods are able to handle an arbitrary number of energy
levels, i.e., the Lindblad equation is solved rather than the (two-level)
optical Bloch equations.

The basic library of mbsolve provides a flexible and extensible framework
to describe devices to be modeled, and simulation scenarios.
On this basis, solvers can be written that implement different numerical
methods and/or parallelization techniques.
Similarly, writers can be implemented that export the simulation results to
certain file formats.
We have implemented and described two solvers that base on the FDTD
method and use the OpenMP standard for parallelization, two algorithms for
solving the Lindblad equation, and a writer for the HDF5 file format.
The resulting source code is written in C++ and features automatically
generated bindings for Python.
It is open-source and can be compiled using most of the established C++
compilers on all major platforms.
Alternatively, the mbsolve software can be installed in binary form using
the conda package manager.

We have demonstrated the usage of our software with the help of four
application examples, selected to cover different use cases.
The first example features a two-level medium that can be described using a
simplified version of the quantum mechanical description.
The setup of the second simulation example can be considered a driven quantum
mechanical system.
In this simulation, there is no spatial coordinate, i.e., field propagation
effects are not considered.
The third example uses the generalized quantum mechanical description to
handle the six-level medium.
Finally, the last example is a simulation of an actual quantum cascade laser.
This example is the most complex and computationally most demanding simulation.

Although the mbsolve software is already a helpful and reliable tool,
several interesting issues are still unsolved, and potential optimization
possibilities remain to be exploited.
The discussion of numerical methods is ongoing and we hope that in the near
future additional methods are implemented in the mbsolve project.
Although the methods for the Lindblad equation seem to have more room for
improvement, there are also promising alternative candidates for Maxwell's
equations (e.g., methods that use adaptive spatial grids).
While we have focused on methods that avoid the RWA/SVEA, we note that
those approximations are beneficial for several applications.
In future work, a solver that uses a numerical method that invokes the RWA
(such as the Risken-Nummedal scheme~\cite{risken1968}) could be implemented
to support these applications better.
As to parallelization, the existing solvers are to be extended to
distributed memory systems using Message Passing Interface (MPI) standard.
Furthermore, offloading of calculations to graphics processing units (GPUs)
is an intriguing possibility.

From the modeling point of view, the inclusion of dispersion in the
electromagnetic properties will be the next step.
While it is not exactly trivial to implement, it is bound to play a
significant role in the modeling of optoelectronic devices (see e.g.,
\cite{singleton2018}).
Also, the implementation of alternative boundary conditions will extend
the application range of mbsolve, e.g., to the simulation of ring cavities.
Finally, in order to enable numerically efficient simulations in two or
three spatial dimensions, an attractive strategy could be to integrate
the mbsolve code for the Lindblad equation into an established and
high-performance open-source electromagnetics simulation project, such as MEEP.

\section*{Acknowledgments}

This work was supported by the German Research Foundation (DFG) within the
Heisenberg program (JI 115/4-2). The authors gratefully acknowledge the help
of their students Mariem Kthiri, Tien Dat Nguyen, Alek Pikl,
Sebastian Senninger, Wenhua Shi, Christian Widmann, and Yi Zhang.
Finally, the authors thank Petar Tzenov for many stimulating discussions.

%% file: figures/tikz_fdtd_os.tex
\begin{tikzpicture}
  \tikzstyle{myline}=[line cap=round,line join=round, ultra thick];
  %
  % coordinate system
  \tikzstyle{cs}=[myline, black, ->, >=stealth];
  \begin{scope}
    \draw[cs] (0, 0) -- (1, 0) node[above] {$x$};
    \draw[cs] (0, 0) -- (0, -1) node[left] {$t$};
  \end{scope}
  %
  % draw magnetic field with blue circles
  \foreach \i in {0, ..., 7} {%
    \foreach \j in {0, ..., 3} {%
      \node (h\i\j) at (0.5 + 1.0*\i, -0.5 - 1.0*\j) { };
      \draw[myline, sipblue] (h\i\j) circle[radius=3pt];
    }
  }
  %
  % draw electric field with orange crosses
  \foreach \i in {0, ..., 6} {%
    \foreach \j in {0, ..., 3} {%
      \node (e\i\j) at (1.0 + 1.0*\i, -1.0 - 1.0*\j) { };
      \draw[myline, siporange] (e\i\j) ++ (-3pt, 0) -- ++(6pt, 0);
      \draw[myline, siporange] (e\i\j) ++ (0, -3pt) -- ++(0, 6pt);
    }
  }
  %
  % draw density matrix with green squares
  \foreach \i in {0, ..., 6} {%
    \foreach \j in {0, ..., 3} {%
      \node (d\i\j) at (1.0 + 1.0*\i, -0.5 - 1.0*\j) { };
      \draw[myline, sipgreen] (d\i\j) ++(-3pt, -3pt) rectangle ++(6pt, 6pt);
    }
  }
  %
  % draw magnetic field update
  \draw[cs] ($(h30) + (0, -5pt)$) -- ($(h31) + (0, 5pt)$);
  \draw[cs] ($(e20) + (3pt, -3pt)$) -- ($(h31) + (0, 5pt)$);
  \draw[cs] ($(e30) + (-3pt, -3pt)$) -- ($(h31) + (0, 5pt)$);
  %
  % draw electric field update
  \draw[cs] ($(e51) + (0, -5pt)$) -- ($(e52) + (0, 5pt)$);
  \draw[cs] ($(h52) + (3pt, -3pt)$) -- ($(e52) + (0, 5pt)$);
  \draw[cs] ($(h62) + (-3pt, -3pt)$) -- ($(e52) + (0, 5pt)$);
  \draw[cs, out=-30, in=60] ($(d52) + (5pt, 0)$) to ($(e52) + (5pt, 0)$);
  %
  % draw density matrix update
  \draw[cs] ($(d12) + (0, -5pt)$) -- ($(d13) + (0, 5pt)$);
  \draw[cs, out=-30, in=60] ($(e12) + (5pt, 0)$) to ($(d13) + (5pt, 0)$);
\end{tikzpicture}

%% file: article_tum.bbl
\begin{thebibliography}{10}
\expandafter\ifx\csname url\endcsname\relax
  \def\url#1{\texttt{#1}}\fi
\expandafter\ifx\csname urlprefix\endcsname\relax\def\urlprefix{URL }\fi
\expandafter\ifx\csname href\endcsname\relax
  \def\href#1#2{#2} \def\path#1{#1}\fi

\bibitem{boyd2008}
R.~W. Boyd, Nonlinear Optics, 3rd Edition, Academic Press, 2008.

\bibitem{bloch1946}
F.~Bloch, Nuclear induction, Phys. Rev. 70 (1946) 460--474.
\newblock \href {https://doi.org/10.1103/PhysRev.70.460}
  {\path{doi:10.1103/PhysRev.70.460}}.

\bibitem{feynman1957}
R.~P. Feynman, F.~L. Vernon, R.~W. Hellwarth, Geometrical representation of the
  {Schrödinger} equation for solving maser problems, J. Appl. Phys. 28~(1)
  (1957) 49--52.
\newblock \href {https://doi.org/10.1063/1.1722572}
  {\path{doi:10.1063/1.1722572}}.

\bibitem{arecchi1965}
F.~T. Arecchi, R.~Bonifacio, Theory of optical maser amplifiers, IEEE J.
  Quantum Electron. 1~(4) (1965) 169--178.
\newblock \href {https://doi.org/10.1109/JQE.1965.1072212}
  {\path{doi:10.1109/JQE.1965.1072212}}.

\bibitem{abella1966}
I.~D. Abella, N.~A. Kurnit, S.~R. Hartmann, Photon echoes, Phys. Rev. 141
  (1966) 391--406.
\newblock \href {https://doi.org/10.1103/PhysRev.141.391}
  {\path{doi:10.1103/PhysRev.141.391}}.

\bibitem{mccall1967}
S.~L. McCall, E.~L. Hahn, Self-induced transparency by pulsed coherent light,
  Phys. Rev. Lett. 18 (1967) 908--911.
\newblock \href {https://doi.org/10.1103/PhysRevLett.18.908}
  {\path{doi:10.1103/PhysRevLett.18.908}}.

\bibitem{mccall1969}
S.~L. McCall, E.~L. Hahn, Self-induced transparency, Phys. Rev. 183 (1969)
  457--485.
\newblock \href {https://doi.org/10.1103/PhysRev.183.457}
  {\path{doi:10.1103/PhysRev.183.457}}.

\bibitem{hioe1981}
F.~T. Hioe, J.~H. Eberly, {$N$}-level coherence vector and higher conservation
  laws in quantum optics and quantum mechanics, Phys. Rev. Lett. 47~(12) (1981)
  838--841.
\newblock \href {https://doi.org/10.1103/PhysRevLett.47.838}
  {\path{doi:10.1103/PhysRevLett.47.838}}.

\bibitem{hau1999}
L.~V. Hau, S.~E. Harris, Z.~Dutton, C.~H. Behroozi, Light speed reduction to 17
  metres per second in an ultracold atomic gas, Nature 397~(6720) (1999)
  594--598.
\newblock \href {https://doi.org/10.1038/17561} {\path{doi:10.1038/17561}}.

\bibitem{liu2001}
C.~Liu, Z.~Dutton, C.~H. Behroozi, L.~V. Hau, Observation of coherent optical
  information storage in an atomic medium using halted light pulses, Nature
  409~(6819) (2001) 490--493.
\newblock \href {https://doi.org/10.1038/35054017}
  {\path{doi:10.1038/35054017}}.

\bibitem{ziolkowski1995}
R.~W. Ziolkowski, J.~M. Arnold, D.~M. Gogny, Ultrafast pulse interactions with
  two-level atoms, Phys. Rev. A 52~(4) (1995) 3082--3094.
\newblock \href {https://doi.org/10.1103/PhysRevA.52.3082}
  {\path{doi:10.1103/PhysRevA.52.3082}}.

\bibitem{cartar2017}
W.~Cartar, J.~M\o{}rk, S.~Hughes, Self-consistent {Maxwell-Bloch} model of
  quantum-dot photonic-crystal-cavity lasers, Phys. Rev. A 96~(2) (2017)
  023859.
\newblock \href {https://doi.org/10.1103/PhysRevA.96.023859}
  {\path{doi:10.1103/PhysRevA.96.023859}}.

\bibitem{gladysz2020}
P.~G{\l}adysz, Piotr~Wcis{\l}o, K.~S{\l}owik, Propagation of optically tunable
  coherent radiation in a medium of asymmetric molecules, Sci. Rep. 10~(1)
  (2020) 17615.
\newblock \href {https://doi.org/10.1038/s41598-020-74569-w}
  {\path{doi:10.1038/s41598-020-74569-w}}.

\bibitem{slavcheva2002}
G.~Slavcheva, J.~M. Arnold, I.~Wallace, R.~W. Ziolkowski, Coupled
  {Maxwell-pseudospin} equations for investigation of self-induced transparency
  effects in a degenerate three-level quantum system in two dimensions:
  {Finite-difference} time-domain study, Phys. Rev. A 66~(6) (2002) 63418.
\newblock \href {https://doi.org/10.1103/PhysRevA.66.063418}
  {\path{doi:10.1103/PhysRevA.66.063418}}.

\bibitem{sukharev2011}
M.~Sukharev, A.~Nitzan, Numerical studies of the interaction of an atomic
  sample with the electromagnetic field in two dimensions, Phys. Rev. A 84~(4)
  (2011) 043802.
\newblock \href {https://doi.org/10.1103/PhysRevA.84.043802}
  {\path{doi:10.1103/PhysRevA.84.043802}}.

\bibitem{marskar2011}
R.~Marskar, U.~{\"O}sterberg, Multilevel {Maxwell-Bloch} simulations in
  inhomogeneously broadened media, Opt. Express 19~(18) (2011) 16784--16796.
\newblock \href {https://doi.org/10.1364/OE.19.016784}
  {\path{doi:10.1364/OE.19.016784}}.

\bibitem{jirauschek2019ats}
C.~Jirauschek, M.~Riesch, P.~Tzenov, Optoelectronic device simulations based on
  macroscopic {Maxwell--Bloch} equations, Adv. Theor. Simul. 2~(8) (2019)
  1900018.
\newblock \href {https://doi.org/10.1002/adts.201900018}
  {\path{doi:10.1002/adts.201900018}}.

\bibitem{kazarinov1971}
R.~F. Kazarinov, R.~A. Suris, Possibility of the amplification of
  electromagnetic waves in a semiconductor with a superlattice, Sov. Phys.
  Semicond. 5~(4) (1971) 797--800.

\bibitem{faist1994}
J.~Faist, F.~Capasso, D.~L. Sivco, C.~Sirtori, A.~L. Hutchinson, A.~Y. Cho,
  Quantum cascade laser, Science 264~(5158) (1994) 553--556.
\newblock \href {https://doi.org/10.1126/science.264.5158.553}
  {\path{doi:10.1126/science.264.5158.553}}.

\bibitem{wang2007}
C.~Y. Wang, L.~Diehl, A.~Gordon, C.~Jirauschek, F.~X. K\"artner, A.~Belyanin,
  D.~Bour, S.~Corzine, G.~H\"ofler, M.~Troccoli, J.~Faist, F.~Capasso, Coherent
  instabilities in a semiconductor laser with fast gain recovery, Phys. Rev. A
  75~(3) (2007) 031802.
\newblock \href {https://doi.org/10.1103/PhysRevA.75.031802}
  {\path{doi:10.1103/PhysRevA.75.031802}}.

\bibitem{menyuk2009}
C.~R. Menyuk, M.~A. Talukder, Self-induced transparency modelocking of quantum
  cascade lasers, Phys. Rev. Lett. 102~(2) (2009) 023903.
\newblock \href {https://doi.org/10.1103/PhysRevLett.102.023903}
  {\path{doi:10.1103/PhysRevLett.102.023903}}.

\bibitem{choi2010}
H.~Choi, V.-M. Gkortsas, L.~Diehl, D.~Bour, S.~Corzine, J.~Zhu, G.~H{\"o}fler,
  F.~Capasso, F.~X. K{\"a}rtner, T.~B. Norris, Ultrafast {Rabi} flopping and
  coherent pulse propagation in a quantum cascade laser, Nat. Photonics 4~(10)
  (2010) 706--710.
\newblock \href {https://doi.org/10.1038/nphoton.2010.205}
  {\path{doi:10.1038/nphoton.2010.205}}.

\bibitem{gkortsas2010}
V.-M. Gkortsas, C.~Y. Wang, L.~Kuznetsova, L.~Diehl, A.~Gordon, C.~Jirauschek,
  M.~A. Belkin, A.~Belyanin, F.~Capasso, F.~X. K{\"a}rtner, Dynamics of
  actively mode-locked quantum cascade lasers, Opt. Express 18~(13) (2010)
  13616--13630.
\newblock \href {https://doi.org/10.1364/OE.18.013616}
  {\path{doi:10.1364/OE.18.013616}}.

\bibitem{freeman2013}
J.~R. Freeman, J.~Maysonnave, S.~Khanna, E.~H. Linfield, A.~G. Davies,
  S.~Dhillon, J.~Tignon, Laser-seeding dynamics with few-cycle pulses:
  {Maxwell-Bloch} finite-difference time-domain simulations of terahertz
  quantum cascade lasers, Phys. Rev. A 87~(6) (2013) 063817.
\newblock \href {https://doi.org/10.1103/PhysRevA.87.063817}
  {\path{doi:10.1103/PhysRevA.87.063817}}.

\bibitem{jirauschek2014}
C.~Jirauschek, T.~Kubis, Modeling techniques for quantum cascade lasers, Appl.
  Phys. Rev. 1 (2014) 011307.
\newblock \href {https://doi.org/10.1063/1.4863665}
  {\path{doi:10.1063/1.4863665}}.

\bibitem{talukder2014}
M.~A. Talukder, C.~R. Menyuk, Quantum coherent saturable absorption for
  mid-infrared ultra-short pulses, Opt. Express 22~(13) (2014) 15608--15617.
\newblock \href {https://doi.org/10.1364/OE.22.015608}
  {\path{doi:10.1364/OE.22.015608}}.

\bibitem{wang2015active}
Y.~Wang, A.~Belyanin, Active mode-locking of mid-infrared quantum cascade
  lasers with short gain recovery time, Opt. Express 23~(4) (2015) 4173--4185.
\newblock \href {https://doi.org/10.1364/OE.23.004173}
  {\path{doi:10.1364/OE.23.004173}}.

\bibitem{tzenov2016}
P.~Tzenov, D.~Burghoff, Q.~Hu, C.~Jirauschek, Time domain modeling of terahertz
  quantum cascade lasers for frequency comb generation, Opt. Express 24~(20)
  (2016) 23232--23247.
\newblock \href {https://doi.org/10.1364/OE.24.023232}
  {\path{doi:10.1364/OE.24.023232}}.

\bibitem{vukovic2017}
N.~N. Vukovi{\'c}, J.~Radovanovi{\'c}, V.~Milanovi{\'c}, D.~L. Boiko,
  Low-threshold {RNGH} instabilities in quantum cascade lasers, IEEE J. Sel.
  Top. Quant. 23~(6) (2017) 1--16.
\newblock \href {https://doi.org/10.1109/JSTQE.2017.2699139}
  {\path{doi:10.1109/JSTQE.2017.2699139}}.

\bibitem{tzenov2018}
P.~Tzenov, I.~Babushkin, R.~Arkhipov, M.~Arkhipov, N.~N. Rosanov, U.~Morgner,
  C.~Jirauschek, Passive and hybrid mode locking in multi-section terahertz
  quantum cascade lasers, New J. Phys. 20~(5) (2018) 053055.
\newblock \href {https://doi.org/10.1088/1367-2630/aac12a}
  {\path{doi:10.1088/1367-2630/aac12a}}.

\bibitem{columbo2018}
L.~Columbo, S.~Barbieri, C.~Sirtori, M.~Brambilla, Dynamics of a broad-band
  quantum cascade laser: from chaos to coherent dynamics and mode-locking, Opt.
  Express 26~(3) (2018) 2829--2847.
\newblock \href {https://doi.org/10.1364/OE.26.002829}
  {\path{doi:10.1364/OE.26.002829}}.

\bibitem{nielsen2007}
P.~K. Nielsen, H.~Thyrrestrup, J.~M{\o}rk, B.~Tromborg, Numerical investigation
  of electromagnetically induced transparency in a quantum dot structure, Opt.
  Express 15~(10) (2007) 6396--6408.
\newblock \href {https://doi.org/10.1364/OE.15.006396}
  {\path{doi:10.1364/OE.15.006396}}.

\bibitem{majer2010}
N.~Majer, K.~L{\"u}dge, E.~Sch{\"o}ll, Cascading enables ultrafast gain
  recovery dynamics of quantum dot semiconductor optical amplifiers, Phys. Rev.
  B 82~(23) (2010) 235301.
\newblock \href {https://doi.org/10.1103/PhysRevB.82.235301}
  {\path{doi:10.1103/PhysRevB.82.235301}}.

\bibitem{slavcheva2019}
G.~Slavcheva, M.~Koleva, A.~Rastelli, Ultrafast pulse phase shifts in a
  charged-quantum-dot--micropillar system, Phys. Rev. B 99~(11) (2019) 115433.
\newblock \href {https://doi.org/10.1103/PhysRevB.99.115433}
  {\path{doi:10.1103/PhysRevB.99.115433}}.

\bibitem{riesch2018atrasc1}
M.~Riesch, P.~Tzenov, C.~Jirauschek, Dynamic simulations of quantum cascade
  lasers beyond the rotating wave approximation, in: 2018 2nd URSI Atlantic
  Radio Science Meeting (AT-RASC), IEEE, Piscataway, NJ, 2018.
\newblock \href {https://doi.org/10.23919/URSI-AT-RASC.2018.8471596}
  {\path{doi:10.23919/URSI-AT-RASC.2018.8471596}}.

\bibitem{riesch2018iqclsw}
M.~Riesch, J.~H. Abundis-Patino, P.~Tzenov, C.~Jirauschek, Efficient simulation
  of the quantum cascade laser dynamics beyond the rotating wave approximation,
  in: {Proceedings of the International Quantum Cascade Laser School and
  Workshop (IQCLSW) 2018}, 2018.

\bibitem{bidegaray2001}
B.~Bid{\'e}garay, A.~Bourgeade, D.~Reignier, Introducing physical relaxation
  terms in {Bloch} equations, J. Comput. Phys. 170~(2) (2001) 603--613.

\bibitem{bidegaray2003}
B.~Bid{\'e}garay, Time discretizations for {Maxwell-Bloch} equations, Numer.
  Methods Partial Differ. Equ. 19~(3) (2003) 284--300.
\newblock \href {https://doi.org/10.1002/num.10046}
  {\path{doi:10.1002/num.10046}}.

\bibitem{saut2006}
O.~Saut, A.~Bourgeade, Numerical methods for the bidimensional {Maxwell--Bloch}
  equations in nonlinear crystals, J. Comput. Phys. 213~(2) (2006) 823--843.
\newblock \href {https://doi.org/10.1016/j.jcp.2005.09.003}
  {\path{doi:10.1016/j.jcp.2005.09.003}}.

\bibitem{deinega2014}
A.~Deinega, T.~Seideman, Self-interaction-free approaches for self-consistent
  solution of the {Maxwell-Liouville} equations, Phys. Rev. A 89~(2) (2014)
  022501.

\bibitem{riesch2018oqel}
M.~Riesch, N.~Tchipev, S.~Senninger, H.-J. Bungartz, C.~Jirauschek, Performance
  evaluation of numerical methods for the {Maxwell--Liouville--von Neumann}
  equations, Opt. Quant. Electron. 50~(2) (2018) 112.
\newblock \href {https://doi.org/10.1007/s11082-018-1377-4}
  {\path{doi:10.1007/s11082-018-1377-4}}.

\bibitem{riesch2019cleo1}
M.~Riesch, C.~Jirauschek, Completely positive trace preserving numerical
  methods for long-term generalized {Maxwell-Bloch} simulations, in: {Lasers
  and Electro-Optics Europe \& European Quantum Electronics Conference
  (CLEO/Europe-EQEC), 2019 Conference on}, IEEE, Piscataway, NJ, 2019.
\newblock \href {https://doi.org/10.1109/CLEOE-EQEC.2019.8873263}
  {\path{doi:10.1109/CLEOE-EQEC.2019.8873263}}.

\bibitem{riesch2020nusod1}
M.~Riesch, A.~Pikl, C.~Jirauschek, Completely positive trace preserving methods
  for the {Lindblad} equation, in: Numerical Simulation of Optoelectronic
  Devices (NUSOD), 2020 International Conference on, IEEE, Piscataway, NJ,
  2020.
\newblock \href {https://doi.org/10.1109/NUSOD49422.2020.9217670}
  {\path{doi:10.1109/NUSOD49422.2020.9217670}}.

\bibitem{emtl}
Kintechlab, Electromagnetic template library,
  \url{http://fdtd.kintechlab.com/en/start} (2018).

\bibitem{freetwm}
J.~Javaloyes, S.~Balle, Freetwm: a simulation tool for semiconductor lasers,
  \url{https://onl.uib.eu/Softwares/Freetwm/} (2018).

\bibitem{oskooi2010}
A.~F. Oskooi, D.~Roundy, M.~Ibanescu, P.~Bermel, J.~D. Joannopoulos, S.~G.
  Johnson, Meep: A flexible free-software package for electromagnetic
  simulations by the {FDTD} method, Comput. Phys. Commun. 181~(3) (2010)
  687--702.
\newblock \href {https://doi.org/10.1016/j.cpc.2009.11.008}
  {\path{doi:10.1016/j.cpc.2009.11.008}}.

\bibitem{riesch2019pasc}
M.~Riesch, N.~Tchipev, H.-J. Bungartz, C.~Jirauschek, Numerical simulation of
  the quantum cascade laser dynamics on parallel architectures, in: Proceedings
  of the Platform for Advanced Scientific Computing Conference, ACM, New York,
  NY, 2019, pp. 5:1--5:8.
\newblock \href {https://doi.org/10.1145/3324989.3325715}
  {\path{doi:10.1145/3324989.3325715}}.

\bibitem{wang2020ultrafast}
F.~Wang, V.~Pistore, M.~Riesch, H.~Nong, P.-B. Vigneron, R.~Colombelli,
  O.~Parillaud, J.~Mangeney, J.~Tignon, C.~Jirauschek, S.~S. Dhillon, Ultrafast
  response of active and self-starting harmonic modelocked {THz} laser, Light
  Sci. Appl. 9~(1) (2020) 51.
\newblock \href {https://doi.org/10.1038/s41377-020-0288-x}
  {\path{doi:10.1038/s41377-020-0288-x}}.

\bibitem{mezzapesa2020}
F.~P. Mezzapesa, K.~Garrasi, J.~Schmidt, L.~Salemi, V.~Pistore, L.~Li, A.~G.
  Davies, E.~H. Linfield, M.~Riesch, C.~Jirauschek, T.~Carey, F.~Torrisi, A.~C.
  Ferrari, M.~S. Vitiello, Terahertz frequency combs exploiting an on-chip,
  solution processed, graphene-quantum cascade laser coupled-cavity, ACS
  Photonics 7~(12) (2020) 3489--3498.
\newblock \href {https://doi.org/10.1021/acsphotonics.0c01523}
  {\path{doi:10.1021/acsphotonics.0c01523}}.

\bibitem{siegman1986}
A.~E. Siegman, Lasers, University Science Books, Mill Valley, California, 1986.

\bibitem{song2005}
X.~Song, S.~Gong, Z.~Xu, Propagation of a few-cycle laser pulse in a {V}-type
  three-level system, Opt. Spectrosc. 99~(4) (2005) 517--521.
\newblock \href {https://doi.org/10.1134/1.2113361}
  {\path{doi:10.1134/1.2113361}}.

\bibitem{taflove2005}
A.~Taflove, S.~C. Hagness, {Computational Electrodynamics: The
  Finite-Difference Time-Domain Method}, Artech House, 2005.

\bibitem{riesch2019jcomp}
M.~Riesch, C.~Jirauschek, Analyzing the positivity preservation of numerical
  methods for the {Liouville-von Neumann} equation, J. Comput. Phys. 390 (2019)
  290--296.
\newblock \href {https://doi.org/10.1016/j.jcp.2019.04.006}
  {\path{doi:10.1016/j.jcp.2019.04.006}}.

\bibitem{riesch2017c}
M.~Riesch, C.~Jirauschek, Numerical method for the
  {Maxwell-Liouville-von Neumann} equations using efficient matrix exponential
  computations,
  \href{https://arxiv.org/abs/1710.09799}{\path{arxiv:1710.09799}}.

\bibitem{riesch2017cleo}
M.~Riesch, N.~Tchipev, H.-J. Bungartz, C.~Jirauschek, Solving the
  {Maxwell-Bloch} equations efficiently on parallel architectures, in: {Lasers
  and Electro-Optics Europe \& European Quantum Electronics Conference
  (CLEO/Europe-EQEC), 2017 Conference on}, IEEE, Piscataway, NJ, 2017.
\newblock \href {https://doi.org/10.1109/CLEOE-EQEC.2017.8087734}
  {\path{doi:10.1109/CLEOE-EQEC.2017.8087734}}.

\bibitem{riesch2020pone}
M.~Riesch, T.~D. Nguyen, C.~Jirauschek, {bertha:} {Project} skeleton for
  scientific software, PLOS ONE 15~(3) (2020) e0230557.
\newblock \href {https://doi.org/10.1371/journal.pone.0230557}
  {\path{doi:10.1371/journal.pone.0230557}}.

\bibitem{qclsip}
M.~Riesch, {qclsip: The Quantum Cascade Laser Stock Image Project} (Jan. 2019).
\newblock \href {https://doi.org/10.5281/zenodo.2641239}
  {\path{doi:10.5281/zenodo.2641239}}.

\bibitem{breuer2002}
H.-P. Breuer, F.~Petruccione, The Theory of Open Quantum Systems, Oxford
  University Press, Oxford, 2002.

\bibitem{lindblad1976}
G.~Lindblad, On the generators of quantum dynamical semigroups, Commun. Math.
  Phys. 48~(2) (1976) 119--130.
\newblock \href {https://doi.org/10.1007/BF01608499}
  {\path{doi:10.1007/BF01608499}}.

\bibitem{gorini1976}
V.~Gorini, A.~Kossakowski, E.~C.~G. Sudarshan, Completely positive dynamical
  semigroups of {$N$}-level systems, J. Math. Phys. 17~(5) (1976) 821--825.
\newblock \href {https://doi.org/10.1063/1.522979}
  {\path{doi:10.1063/1.522979}}.

\bibitem{schirmer2004}
S.~G. Schirmer, A.~I. Solomon, Constraints on relaxation rates for {$N$}-level
  quantum systems, Phys. Rev. A 70 (2004) 022107.
\newblock \href {https://doi.org/10.1103/PhysRevA.70.022107}
  {\path{doi:10.1103/PhysRevA.70.022107}}.

\bibitem{oi2012}
D.~K.~L. Oi, S.~G. Schirmer, Limits on the decay rate of quantum coherence and
  correlation, Phys. Rev. A 86 (2012) 012121.
\newblock \href {https://doi.org/10.1103/PhysRevA.86.012121}
  {\path{doi:10.1103/PhysRevA.86.012121}}.

\bibitem{mbsolve-doc}
mbsolve documentation on {GitHub} {Pages},
  \url{https://mriesch-tum.github.io/mbsolve} (2020).

\bibitem{mbsolve-github}
M.~Riesch, C.~Jirauschek, {mbsolve}: {An} open-source solver tool for the
  {Maxwell-Bloch} equations, \url{https://github.com/mriesch-tum/mbsolve} (Jul.
  2017).

\bibitem{burghoff2014}
D.~Burghoff, T.-Y. Kao, N.~Han, C.~W.~I. Chan, X.~Cai, Y.~Yang, D.~J. Hayton,
  J.-R. Gao, J.~L. Reno, Q.~Hu, Terahertz laser frequency combs, Nature Photon.
  8~(6) (2014) 462--467.
\newblock \href {https://doi.org/10.1038/nphoton.2014.85}
  {\path{doi:10.1038/nphoton.2014.85}}.

\bibitem{risken1968}
H.~Risken, K.~Nummedal, Self-pulsing in lasers, J. Appl. Phys. 39~(10) (1968)
  4662--4672.
\newblock \href {https://doi.org/10.1063/1.1655817}
  {\path{doi:10.1063/1.1655817}}.

\bibitem{singleton2018}
M.~Singleton, P.~Jouy, M.~Beck, J.~Faist, Evidence of linear chirp in
  mid-infrared quantum cascade lasers, Optica 5~(8) (2018) 948--953.
\newblock \href {https://doi.org/10.1364/OPTICA.5.000948}
  {\path{doi:10.1364/OPTICA.5.000948}}.

\end{thebibliography}
